\documentclass[letterpaper]{article}
\pdfoutput=1
\usepackage{jcappub}
\usepackage{amsmath}
\usepackage{graphicx}
\usepackage{feynmf}
\usepackage{dsfont}
\usepackage{array}
\usepackage{slashed}
\usepackage{afterpage}
\usepackage{appendix}
\usepackage{multirow}
\usepackage{wrapfig}
\usepackage{cancel}
\usepackage{xcolor}
\usepackage{hyperref}
\usepackage{comment}
\usepackage{siunitx}
\usepackage[T1]{fontenc}
\newcommand{\draftnoteYZ}[1]{\textcolor{cyan}{\textbf{YZ : #1}}}

\title{
BBN Constraint on Heavy Neutrino Production and Decay
}
\author{Yu-Ming Chen,}
\author{and Yue Zhang\,}
\affiliation{Department of Physics, Carleton University, Ottawa, Ontario K1S 5B6, Canada}

\emailAdd{yumingchen@cmail.carleton.ca}
\emailAdd{yzhang@physics.carleton.ca}

\abstract{
We explore the big-bang nucleosynthesis (BBN) constraint on heavy neutrino that is a mixture of gauge singlet fermion and active neutrinos in the Standard Model. We work in the minimal model with only two parameters, the heavy neutrino mass $m_4$ and the mixing parameter $|U_{a4}|^2$, where $a=e$, $\mu$, or $\tau$ stands for the active neutrino flavor. We show that both the early universe production mechanism and decay products of the heavy neutrino are determined by $m_4$ and $|U_{a4}|^2$, with little room for further assumptions. This predictability allows us to present a portrait of the entire BBN excluded parameter space. Our analysis includes various effects including temporary matter domination, energy injections in the form of charged mesons, photons and light neutrinos. The BBN constraint is complementary to terrestrial search for heavy neutrinos (heavy neutral leptons) behind the origin of neutrino masses and portal to the dark sector.
}


\arxivnumber{}

\begin{document}

\maketitle

\section{Introduction}

The neutrino sector of the Standard Model hosts a number of opportunities for new physics searches. The phenomena of neutrino oscillations reveal that neutrinos have nonzero masses and mixings, which require beyond the Standard Model physics~\cite{Super-Kamiokande:1998kpq, SNO:2001kpb, SNO:2002tuh}. Through the type-I seesaw mechanism~\cite{Minkowski:1977sc, Yanagida:1979as, Glashow:1979nm, Mohapatra:1979ia, 
Gell-Mann:1979vob}, neutrinos obtain their masses by coupling to gauge-singlet sterile neutrinos that are Majorana fermions. This scenario could be tested at low energies in lepton number violating processes such as neutrinoless double beta decay~\cite{Furry:1939qr, Bilenky:2012qi, Engel:2016xgb, Dolinski:2019nrj, Dekens:2020ttz, Bolton:2022tds}, and by hunting the heavy neutrino as novel, elusive particles using terrestrial and cosmological experiments~\cite{Friedberg:1963zz, Chang:1968zz, Keung:1983uu, Atre:2009rg, Nemevsek:2011hz, Mitra:2011qr, Drewes:2013gca, Deppisch:2015qwa, Maiezza:2015lza, deGouvea:2015euy, Cai:2017mow, Zhang:2022cyr, Nemevsek:2022anh, Bose:2022obr, Bi:2024pkk, Wang:2024prt, Ajmal:2024kwi}. Besides neutrino mass generation models, heavy right-handed neutrinos (as Majorana or Dirac fermions) are often considered as a portal to dark sectors which accommodate dark matter particles that fill the universe~\cite{Bertoni:2014mva, Berryman:2017twh, Batell:2017cmf, Orlofsky:2021mmy, Zhang:2023mcv}.

The big-bang nucleosynthesis (BBN) provides a unique lens into the very early universe for exploring the Standard Model and beyond~\cite{1967ApJ...148....3W, Kawano:1992ua, Serpico:2004gx, Pospelov:2006sc}. The fact that BBN predictions of primordial element abundances agree well with the observations in standard cosmology limits the room of new physics, including the equation of state and expansion rate of the universe, as well as exotic energy injections. These effects could occur if a population of heavy neutrinos decays into Standard Model particles during the BBN epoch~\cite{Barbieri:1990vx,Kainulainen:1990ds,Enqvist:1990ad,Babu:1991at,Dodelson:1993je}. In a number of recent studies, the BBN constraints are presented as an upper bound on the heavy neutrino's lifetime, or a lower bound on the active-sterile neutrino mixing, by assuming that the heavy neutrino decouples from thermal equilibrium in early universe~\cite{Gorbunov:2007ak,Boyarsky:2009ix,Ruchayskiy:2012si, Alekhin:2015byh, Boyarsky:2020dzc, Sabti:2020yrt}.

In this article, we take minimality as the guiding principle and explore simplified models where the mixing of sterile neutrino with the Standard Model active neutrino is the only portal for it to interact with other known particles. Under this assumption, there is little ambiguity about the fate of heavy neutrino in the early universe.
We consider the mixing involves one active neutrino flavor in each case, so that the production and decay of heavy neutrino is essentially dictated by two parameters, the heavy neutrino's mass and its mixing with the active neutrino.

The interplay between the heavy neutrino lifetime and its thermal contact with the Standard Model plasma is a useful guideline for our study.
Note that only sufficiently weakly-coupled heavy neutrino with a small mixing parameter can have a lifetime comparable to the BBN time scale. For sufficiently small mixing, the heavy neutrino may not establish thermal equilibrium with the Standard Model particles. To set the most conservative limit, we assume it begins with zero initial population after inflation and any abundance must be built up through the mixing with active neutrinos. The BBN constraint on heavy neutrino can be evaded if the mixing parameter is tiny so that the heavy neutrino is rarely produced, or if the mixing is so large that the heavy neutrino decays away well before the onset of BBN. As a result, the BBN constraint only applies to a window of intermediate mixing values.
The goal of this work is to quantify the position of this window as a function of the heavy neutrino mass.

\begin{figure}
    \centering
    \includegraphics[width = 0.9\linewidth]{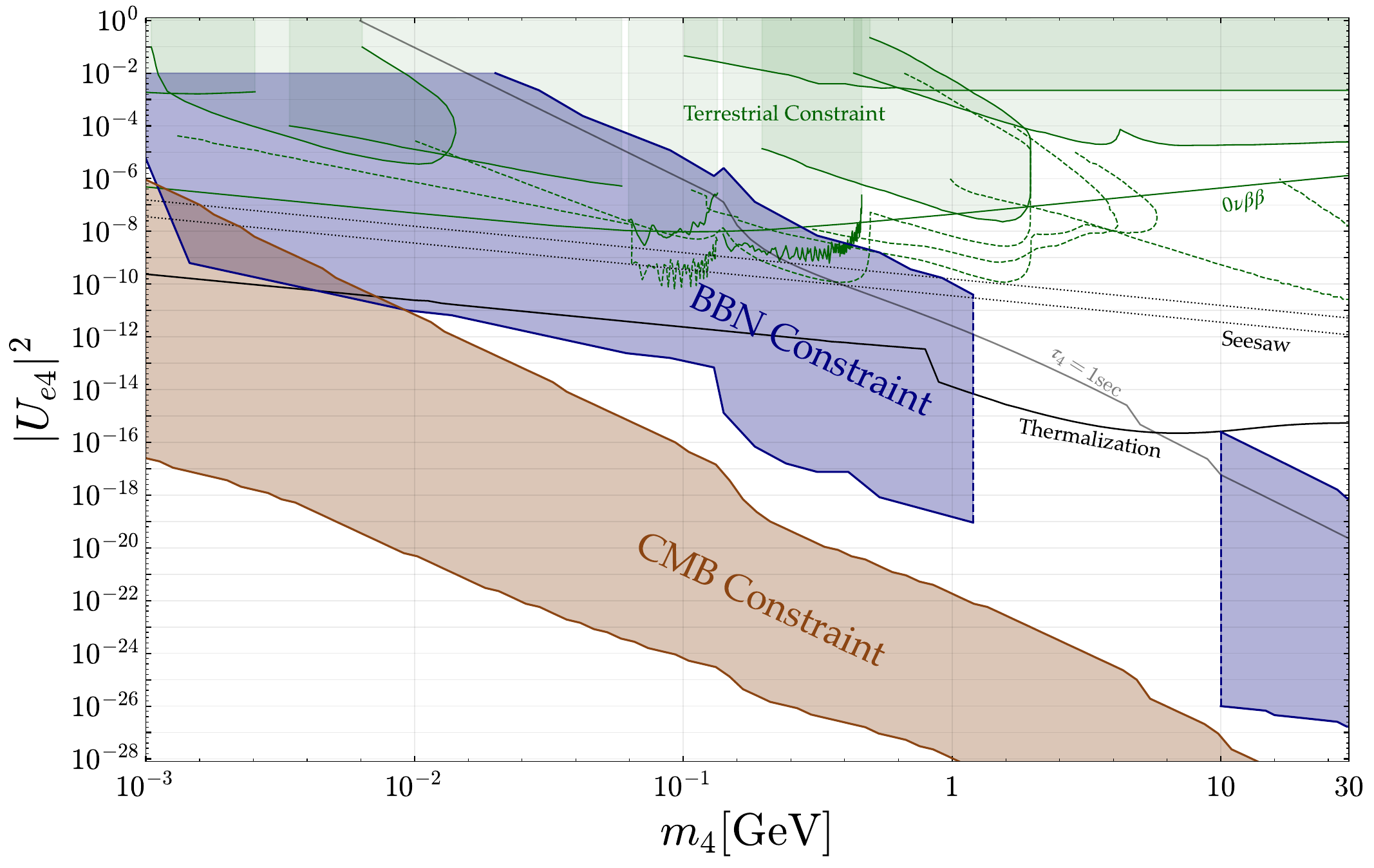}
    \caption{BBN (blue shaded region) and CMB (brown shaded region) constraints derived in this work on the parameter space of heavy neutrino mass versus its mixing with the active neutrino (electron flavor in this case). For BBN, the gap of $m_4$ between a few and 10\,GeV is not shaded with blue where our computation is limited by the complexity of QCD.
    The black dotted lines show the mixing angles when heavy neutrinos explain the active neutrino mass via the seesaw mechanism described in Eq.~(\ref{eq: seesaw}). In the region above the black solid curve, the heavy neutrino can reach thermal equilibrium with the Standard Model plasma in the early universe (see Fig.~\ref{Fig:FOFI}).
    We also show other laboratory constraints for comparison, including the current (green shaded region) and future (green dashed line) limits.
    }
    \label{Fig:MainPlotE}    
\end{figure}

To set the stage, we introduce the minimal model of heavy neutrino that contains a mixture of the active neutrino component. The heavy neutrino mass eigenstate is a linear combination of a gauge singlet fermion $\nu_s$ (sterile neutrino) and active neutrino $\nu_a$. In the vacuum,
\begin{equation}\label{eq:DefMixing}
\nu_4 \simeq \nu_s + U_{a 4} \nu_a \ ,
\end{equation}
where $U_{a4}$ is a matrix element of the neutrino mixing matrix, and $|U_{a4}|$ is much smaller than unity due to existing constraints (see below).
Both $\nu_s$ and $\nu_a$ are flavor eigenstates that can be distinguished by weak interactions. In general, the $\nu_a$ here could be a linear combination of three active neutrino flavors. In this work, we consider a singlet flavor for simplicity in each analysis. The mass of $\nu_4$ we are interested in lies in the range of MeV to the weak scale. In contract, the orthogonal mass eigenstates $\nu_{1,2,3}$ are constrained to be lighter than $\sim 0.1$\,eV~\cite{Planck:2018vyg, DESI:2024mwx}.

Through the above mixing, the mass eigenstate $\nu_4$ can participate in weak interactions mediated by the $W, Z$ bosons, with couplings suppressed by the small $U_{a4}$,
\begin{equation}
\mathcal{L}_{\rm int} = \frac{e U_{a 4}}{\sqrt 2\sin\theta_w} \bar \ell_a \gamma^\mu \mathbb{P}_L \nu_4 W^-_\mu  + \frac{e U_{a 4}}{\sin(2\theta_w)} \bar \nu_a \gamma^\mu \mathbb{P}_L \nu_4 Z_\mu + {\rm h.c.} \ ,
\end{equation}
where $\theta_w$ is the Weinberg angle and $e$ is the electric charge unit.
These interactions are important for studying the decay of heavy neutrinos.

Taking the type-I seesaw mechanism~\cite{Minkowski:1977sc, Mohapatra:1979ia} as a concrete example, the mixing in Eq.~\eqref{eq:DefMixing} can be generated by the following Lagrangian
\begin{equation}
\mathcal{L}_{\rm seesaw} =\frac{1}{2}\left(\bar N i\cancel{\partial} N + M \bar N N \right) + \left( y \bar L_a H \nu_s + {\rm h.c.} \right) \ ,
\end{equation}
where $N = \nu_s + \nu_s^c$, and $\nu_s = \mathbb{P}_R N$ is the right-handed neutrino. $L_a$ and $H$ are the lepton and Higgs doublets. After electroweak symmetry breaking, $\langle H\rangle = v/\sqrt2$, the active-sterile neutrino mixing parameter is given by
\begin{equation}
U_{a 4} \simeq \frac{y v}{\sqrt{2} M} \simeq \sqrt{\frac{m^\nu_{\rm light}}{M}} \ ,
\label{eq: seesaw}
\end{equation}
and the mass eigenvalue for the heavy, mostly sterile state is $m_4 \simeq M$. 
On the other hand, in neutrino portal dark sector models, the mixing parameter $|U_{a4}|$ is not necessarily tied to the neutrino mass and allowed to take a broader range of values.

Fig.~\ref{Fig:MainPlotE} presents the main results of this work, for heavy neutrino mixing with $\nu_e$. (Results for mixing with $\nu_\mu, \nu_\tau$ are presented in Fig.~\ref{Fig:MainPlotMT}.) The two black dotted lines indicate the typical active-sterile neutrino mixing values in the type-I seesaw mechanism for generating the atmospheric and solar neutrino mass differences. The BBN constraints are shown by the blue shaded region. Clearly they have a large overlap with the seesaw target for explaining the light neutrino masses and they are highly complementary to the terrestrial experiments which probe the higher $|U_{a4}|^2$ values (green shaded regions).~\footnote{The present and future reach of terrestrial experiments are taken from \url{https://www.hep.ucl.ac.uk/~pbolton/}. The neutrinoless double beta decay curve is adopted from Ref.~\cite{deGouvea:2015euy}.}
For given heavy neutrino mass $m_4$ (horizontal axis), the BBN exclusion applies for a window of the mixing parameter $|U_{a4}|^2$ (vertical axis). In the parameter space above the blue region, the heavy neutrino decays very early and the effects on BBN are opaqued away by the thermal plasma. In contrast, in the parameter space below the blue region, the heavy neutrino is long-lived and decays during the BBN epoch. However, the mixing parameter is so small that no thermal contact is reached between the heavy neutrino and the Standard Model sector, and the freeze-in abundance is too small to affect the primordial element abundances.
In that region, we show a constraint due to excessive energy injection from the decay of heavy neutrino during the recombination era which distorts the cosmic microwave background (CMB).

\section{Heavy Neutrino Production in the Early Universe}\label{sec:2}

We first address the production of heavy neutrinos in the early universe. The black solid line in Fig.~\ref{Fig:MainPlotE} divides the parameter space into two parts. In the large $|U_{a4}|^2$ region above the line, heavy neutrinos are populated via the regular thermal freeze out mechanism, whereas in the small $|U_{a4}|^2$ region the production is freeze in via neutrino oscillation processes~\cite{Barbieri:1990vx,Kainulainen:1990ds,Enqvist:1990ad,Babu:1991at,Dodelson:1993je}. 

An important player in both production mechanisms is the effective mixing angle between the active and sterile neutrinos. In the early universe plasma, this mixing is modified from its vacuum value in Eq.~\eqref{eq:DefMixing} due to weak interactions acting on the active neutrino state. This manifests as a potential term that alters neutrino's dispersion relation and a scattering rate acting as a damping effect to the oscillation process. Both effects are temperature dependent. 
The active-to-sterile neutrino oscillation problem can be described starting from the quantum kinetic equation of the density matrix~\cite{Sigl:1993ctk}. 
In the limit where the weak interaction has a much larger rate than the expansion of the universe (i.e., $T\gg \,{\rm MeV}$), the set of density matrix equations can be simplified to a simple Boltzmann equation that governs
the phase space distribution (PSD) function of sterile neutrino, 
\begin{equation}\label{eq:PSDBoltzmannEq}
\frac{\partial f_s (x, a)}{\partial \ln a} = \frac{\Gamma \sin^2(2\theta_{\rm eff})}{4H} (f_{\rm eq} - f_s) \ .
\end{equation}
We refer to the appendix~\ref{app:A} for more details.
Here, the argument of the differential equation $a$ is the scale factor of the universes, $H = \dot a/a$ is the Hubble parameter, and
\begin{equation}\label{eq:PSD}
f_{\rm eq}(x) = \frac{1}{1 + e^{x}} \ ,
\end{equation}
is the Fermi-Dirac distribution function, $x= E/T$, and $E$ is the oscillating neutrino energy.
The production of sterile neutrino occurs when it is still ultra-relativistic, thus $x$ is time (temperature, or $a$) independent.
In the early universe plasma, the effective mixing angle between active and sterile neutrino is
\begin{equation}\label{eq:thetaeff}
\sin^2(2\theta_{\rm eff}) = \frac{\Delta^2 \sin^2(2\theta)}{\Delta^2 \sin^2(2\theta) +\left(\Delta\cos(2\theta)-V\right)^2 +  (\Gamma/2)^2} \ .
\end{equation}
Here $\theta$ is the vacuum mixing angle and approximately equals to $|U_{a4}|$ introduced in Eq.~\eqref{eq:DefMixing}, thus
\begin{equation}
\sin(2\theta) \simeq 2 |U_{a4}|\ , \quad \cos(2\theta)\simeq 1 \ . 
\end{equation}

On the right-hand side of Eq.~\eqref{eq:PSDBoltzmannEq}, the quantities $\Delta$, $V$, $\Gamma$ are all functions of $x$ and $a$.
 $\Delta = m_4^2/(2E)$ is the vacuum oscillation frequency, and the thermal potential for active neutrino is~\cite{Notzold:1987ik},
\begin{equation}
\begin{aligned}
V &= - {\rm Re}\sum_{i=W, Z}\frac{g_i^2}{16 \pi^2 E^2} \int_0^\infty dp
\left[
\left(
\frac{M_i^2 p}{2\omega} L_2^+(E,p) - \frac{4Ep^2}{\omega}
\right)
\frac{1}{e^{\omega/T} - 1} +
\left(
\frac{M_i^2}{2} L_1^+(E,p) - 4Ep
\right)
\frac{1}{e^{p/T} + 1}
\right], \\
&L_1^+(E,p) = \ln \frac{4pE + M_i^2}{4pE - M_i^2}, \quad 
    L_2^+(E,p) = \ln \frac{(2pE + 2E \omega + M_i^2)(2pE - 2E\omega + M_i^2)}{(-2pE + 2E \omega + M_i^2)(-2pE - 2E\omega + M_i^2)},
\end{aligned}
\end{equation}
where $\omega = \sqrt{p^2 + M_i^2}$, the index $i$ sums over thermal loop contributions involving the $W, Z$ bosons and the corresponding couplings are 
$g_W=\sqrt{4\sqrt{2} G_F M_W^2}$ and $g_Z = g_W/(\sqrt{2}\cos\theta_W)$. Numerically, $g_W \simeq 0.65$, $g_Z \simeq 0.53$.
For simplicity, we assume zero neutrino chemical potential (asymmetry) throughout this work. In the limit $E, T \ll M_W, M_Z$, the above potential approximates to
\begin{equation}
V = - \frac{7 \sqrt2 \pi^2 G_F E T^4}{45} \left( \frac{2}{M_W^2} + \frac{1}{M_Z^2} \right) \ .
\end{equation}

$\Gamma$ is the total reaction rate for an active neutrino state with energy $E$ and sums over all the processes involving the active neutrino with the ambient particles in the thermal plasma. At temperatures below the weak scale, it takes the form
\begin{equation}\label{eq:GammaA}
    \Gamma = c_\nu G_F^2 E T^4 \ .
\end{equation}
The value of coefficient $c_\nu$ depends on the temperature and the flavor of active neutrino that mixes with the sterile neutrino, and is given in the following table.
At $T<m_e$, the coefficient $c_{\nu_e} = 7\pi/24$ was used in~\cite{Dodelson:1993je}. Going to higher temperatures, the coefficient $c_\nu$ is enhanced by the multiplicity of particle species in the thermal plasma. 
At given $T$, we derive the $c_\nu$ value by neglecting light fermion masses ($m_f<T$) and decoupling the heavy ones.

\begin{equation}\label{eq:GammaCA}
c_\nu = 
\begin{tabular}{| c | c | c | c |}
\hline
\textrm{temperature range} & $\nu_e$ & $\nu_\mu$ & $\nu_\tau$ \\
\hline
\rule{0pt}{15pt}
$T < m_e$ &
$\frac{7 \pi}{27}$ &
$\frac{7 \pi}{27}$ &
$\frac{7 \pi}{27}$ \\ [0.5ex]
\hline
\rule{0pt}{15pt}
$m_e<T<m_\mu$ &
$\frac{91 \pi}{216}$ &
$\frac{7 \pi}{24}$ &
$\frac{7 \pi}{24}$ \\ [0.5ex]
\hline
\rule{0pt}{15pt}
$m_\mu < T < \Lambda_{QCD}$ &
$\frac{77 \pi}{108}$ &
$\frac{77 \pi}{108}$ &
$\frac{35 \pi}{108}$ \\ [0.5ex]
\hline
\rule{0pt}{15pt}
$\Lambda_{QCD} < T < m_c$ & 
$\frac{203 \pi}{108}$ & 
$\frac{203 \pi}{108}$ & 
$\frac{77 \pi}{108}$ \\ [0.5ex]
\hline
\rule{0pt}{15pt}
$m_c < T < m_\tau$ & 
$\frac{244 \pi}{81}$ & 
$\frac{244 \pi}{81}$ & 
$\frac{133 \pi}{162}$ \\ [0.5ex]
\hline
\rule{0pt}{15pt}
$m_\tau < T < m_b$ & 
$\frac{1981 \pi}{648}$ & 
$\frac{1981 \pi}{648}$ & 
$\frac{1981 \pi}{648}$ \\ [0.5ex]
\hline
\rule{0pt}{15pt}
$ m_b<T<M_W $ &
$\frac{259 \pi}{81}$ &
$\frac{259 \pi}{81}$ &
$\frac{259 \pi}{81}$ \\ [0.5ex]
\hline
\end{tabular}
\end{equation}

Based on Eq.~\eqref{eq:PSDBoltzmannEq}, whether the sterile neutrino state could reach thermal equilibrium is dictated by the ratio $\Gamma_a \sin^2(2\theta_{\rm eff})/H$. 
Consider a neutrino with average energy $E\sim T$ in the radiation dominated universe, the weak interaction rate and Hubble have simple scaling with the temperature, $\Gamma \sim G_F^2 T^5$ and $H\sim T^2$. On the other hand, the effective mixing angle $\sin^2(2\theta_{\rm eff})$ has a less trivial temperature dependence. At high temperatures when $\Gamma, V \gg \Delta$, the effective mixing angle is suppressed by weak interactions compared to the vacuum value, and we have
$\sin^2(2\theta_{\rm eff}) \sim \Delta^2|U_{a4}|^2/\Gamma^2 \sim T^{-12}$. As the universe cools, $\Gamma_a$ and $V_a$ decrease whereas $\Delta$ grows; $\theta_{\rm eff}$ eventually settles down to $\theta \simeq |U_{a4}|$. This interplay defines a critical temperature $T_*$, with
\begin{equation}\label{eq:T*}
T_* = \left( \frac{m_4^2}{2 c_\nu G_F^2} \right)^{1/6} \simeq 40\,{\rm GeV} \left( \frac{m_4}{1\,{\rm GeV}} \right)^{1/3} \left( \frac{1}{c_\nu} \right)^{1/6} \ ,
\end{equation}
defined by $\Gamma \simeq \Delta$, where $c_\nu$ is the coefficient given in Eq.~\eqref{eq:GammaCA}.
The ratio $\Gamma \sin^2(2\theta_{\rm eff})/H$ is largest at temperature $T_*$, and suppressed at both very large and very small temperatures,
\begin{equation}
\frac{\Gamma \sin^2(2\theta_{\rm eff})}{H}  \propto \left\{
\begin{array}{cl}
T^{-9}, & \quad T> T_* \\
T^3, & \quad T<T_*
\end{array}
\right.
\end{equation}
The peak at $T_*$ corresponds to the epoch where the sterile neutrino has the most thermal contact with the Standard Model sector. It is worth noting that for $m_4$ below the electroweak scale, $T_*$ is always much larger than the sterile neutrino mass $m_4$.

\subsection{Freeze-out Production}\label{sec:relativisitc}

If $\Gamma \sin^2(2\theta_{\rm eff})/H >1$ at temperature $T_*$, the sterile neutrino state will reach thermal equilibrium with the Standard Model plasma for a period around $T_*$. This requires
\begin{equation}\label{eq:FOcondition}
|U_{a4}|^2 \gtrsim \frac{1}{G_F M_{\rm P} m_4} \simeq 10^{-14} \left(\frac{1\,\rm GeV}{m_4}\right) \ ,
\end{equation}
where $M_{\rm P}=1.2\times10^{19}$\,GeV is the Planck scale.

As the universe cools, it eventually drops out of equilibrium when the ratio $\Gamma \sin^2(2\theta_{\rm eff})/H$ drops to around unity. The freeze out temperature can be estimated to be
\begin{equation}
T_{\rm fo} \simeq \left( G_F^2 M_{\rm P} |U_{a4}|^2
\rule{0mm}{4mm}\right)^{-1/3} \ .
\end{equation}
In practice, we find that sterile neutrino always freezes out relativistically for the range of parameter space constrained by BBN. Indeed, the condition for relativistic freeze out $T_{\rm fo}>m_4$ amounts to
\begin{equation}
|U_{a4}|^2 \lesssim \frac{1}{G_F^2 M_{\rm P} m_4^3} \simeq 10^{-4 } \left(\frac{1\,\rm GeV}{m_4}\right)^3 \ .
\end{equation}
This upper bound is consistent with the freeze out condition Eq.~\eqref{eq:FOcondition} for sterile neutrino mass below the electroweak scale.

\begin{figure}
    \centering
    \includegraphics[width = 0.75\linewidth]{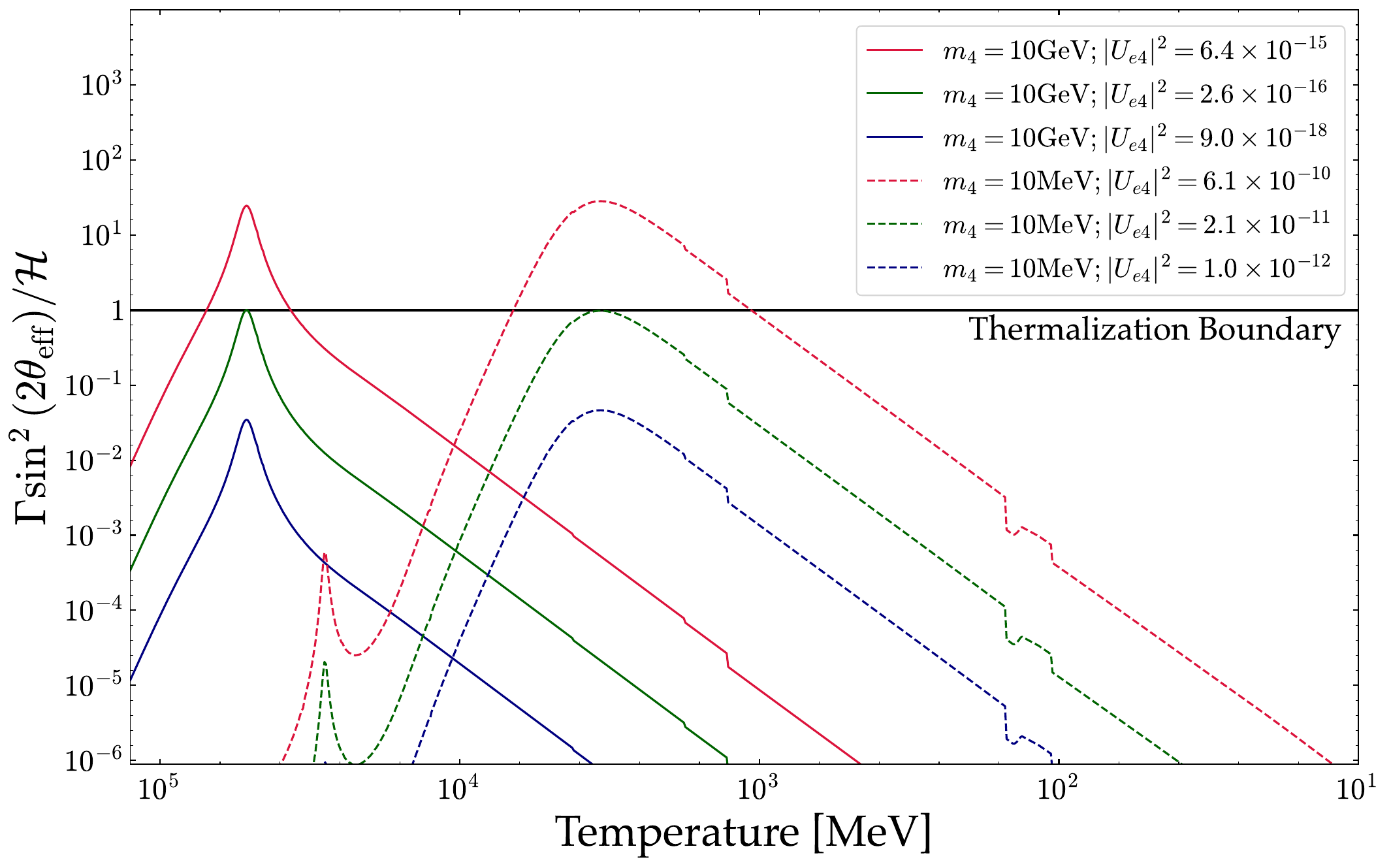}
    \caption{The ratio between the total interaction rate of heavy neutrino in the early universe and the Hubble expansion rate as a function of temperature, for two choices of $m_4$ and various mixing parameter $|U_{\alpha 4}|^2$. Thermalization occurs when the ratio is larger than 1. For both solid and dashed curves, red (blue) corresponds to freeze-out (-in) production, and green shows the marginal case.}
    \label{Fig:FOFI}
\end{figure}

\subsection{Freeze-in Production}\label{sec:FI}

If $\Gamma \sin^2(2\theta_{\rm eff})/H \ll 1$ at temperature $T_*$, the sterile neutrino is never in thermal equilibrium with the Standard Model sector. In this case, its abundance is given by the freeze in mechanism plus an initial condition (assumed to be zero in the minimal model).
We can directly integrate Eq.~\eqref{eq:PSDBoltzmannEq} by dropping the $f_s$ term on the right-hand side
\begin{equation}
\frac{\partial f_s (x, a)}{\partial \ln a} \simeq \frac{\Gamma \sin^2(2\theta_{\rm eff})}{4H} f_{\rm eq} \ .
\end{equation}
The sterile neutrino is dominantly produced at temperature $T_*$. As argued above, $T_*$ is always higher than $m_4$. Therefore, the production occurs in the same way as the Dodelson-Widrow mechanism for keV-scale sterile neutrino dark matter production.
In Fig.~\ref{Fig:FOFI}, we plot $\Gamma \sin^2(2\theta_{\rm eff})/H$ as a function of the temperature (time), for three different value of $|U_{a4}|^2$. The red, blue, green curves corresponds to freeze-out, freeze-in production, and the marginal cases.

A subtle point to clarify is that the sterile neutrino from oscillation is first produced as a pure $\nu_s$ flavor eigenstate (see Eq.~\eqref{eq:PSDBoltzmannEq}), but to discuss decays we must work in the mass basis. In the $|U_{a4}|^2\ll1$ limit, there is no need to distinguish between the two. Nearly 100\% of $\nu_s$ ends up in the heavy neutrino mass eigenstate $\nu_4$.
The probability for $\nu_s$ to oscillate once more is suppressed by another factor of $|U_{a4}|^2$.

After the completion of freezing out or freezing in, the number density of sterile neutrinos simply drops as $a^{-3}$ with the expansion of the universe until decay takes place. 

\subsection{Temporary matter domination}
\label{subsec: Temporary matter domination}

\begin{figure}[t]
    \centering
    \includegraphics[width = 1.0\linewidth]{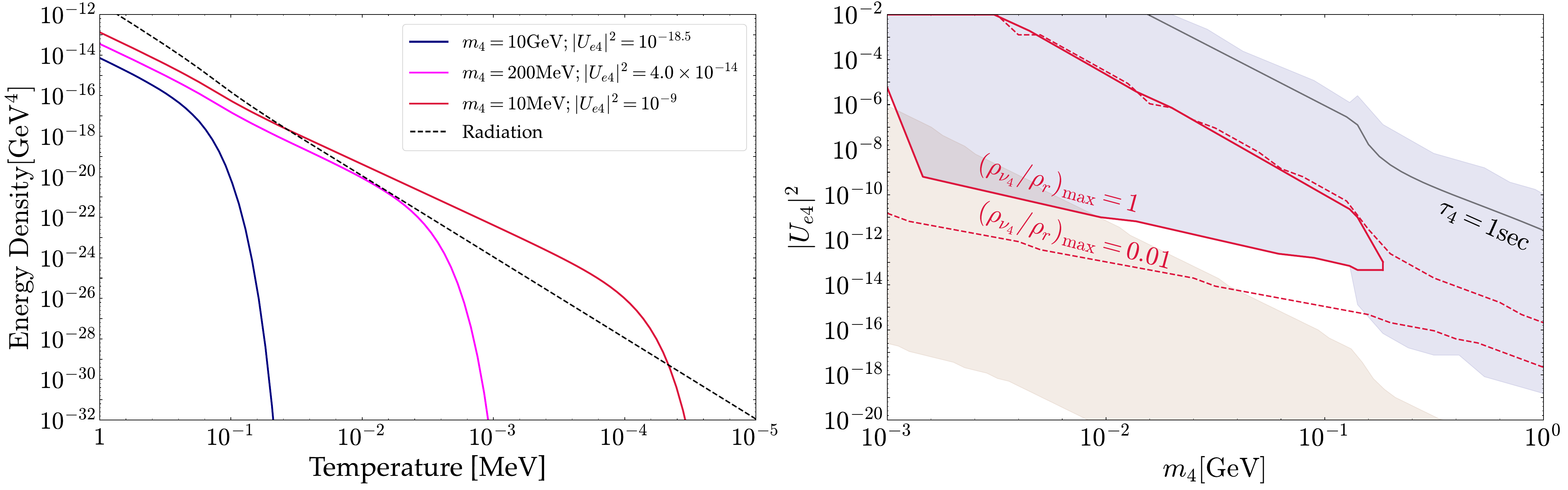}
    \caption{\textbf{Left panel}: Energy density of heavy neutrino as a function of temperature. We show three scenarios where the universe is temporarily dominated by heavy neutrino (red), always radiation dominated (navy), and the marginal case (magenta) The black dashed line is the radiation energy density in the standard cosmology. \textbf{Right panel:} We show the minimal mixing parameter $|U_{a4}|^2$ for the temporary matter domination region (red) and the lifetime equals one second curve (black) on the parameter space of heavy neutrino mass versus mixing. In the background, the light blue and brown shaded regions are BBN and CMB exclusions derived in this work.
    }
    \label{fig:TMDExample}
\end{figure}

The heavy neutrinos made in early universe can impact BBN in several ways. We first discuss an effect prior to the decay taking place. The heavy neutrinos are produced relativistically but later become matter-like as the universe expands. Before decaying away, its energy density redshifts slower than the radiation fluid made of SM particles and could potentially come into temporary domination of the total energy in the universe. 
The left panel of Fig.~\ref{fig:TMDExample} shows the energy density evolution for the heavy neutrino ($\rho_{\nu_4}$) and radiation ($\rho_R$), for three different mass and mixing combinations. 
Temporary matter domination occurs for the red curve.
In our calculation, the Hubble parameter is controlled by the sum of $\rho_{\nu_4}$ and $\rho_R$.

Departure from radiation domination modifies the clock of the universe during BBN, and is tightly constrained~\cite{Yeh:2024ors}.
We require that the heavy neutrino has never dominated the energy density of the universe during BBN, i.e.,
\begin{equation}\label{eq:3.1}
\rho_{\nu_4} < \rho_R \ , \quad \text{for all }\hspace{0.3cm} t\geq 1\,{\rm sec} \ ,
\end{equation}
where 
\begin{equation}
\rho_{\nu_4} = \int \frac{d^3 p}{(2\pi)^3} \sqrt{p^2+m_4^2} f_s \left(\sqrt{p^2+m_4^2}/T\right) \ , \quad
\rho_R = \frac{\pi^2}{30} g_*(T) T^4 \ ,
\end{equation}
and $f_s$ is the PSD function calculated above for the freeze-out or freeze-in mechanisms.
Before the decay taking place, we have $\rho_{\nu_4} \sim a^{4(3)}$ for $T\gg (\ll)\, m_4$, where $a$ is the scale factor of the universe.
Obeying Eq.~\eqref{eq:3.1} ensures that BBN occurs in an entirely radiation dominated universe. 

In the right panel of Fig.~\ref{fig:TMDExample}, the red solid curve shows the mixing parameter $|U_{a4}|^2$ allowing for $(\rho_{\nu_4}/\rho_R)_{\rm max}=1$. In the region surrounded by this solid red curve, temporary matter domination occurs during BBN and is excluded. In the background, the light blue and brown shaded regions are BBN and CMB exclusions derived in this work.
Below the solid red curve, the heavy neutrino is always a subdominant component of the universe and there is no significant entropy dilution. 
Note there exists an unexcluded region immediately below the temporary-matter-domination region with heavy neutrino mass $m_4$ between 10--100 MeV. 
In that region, we show the red dashed curve where the heavy neutrino can reach 1\% of the total energy in the universe before decaying away.

\section{Heavy Neutrino Decay During BBN}\label{sec:nu4Decay}

For the range of parameters considered in this work, the heavy neutrinos are always produced relativistically in both the freeze out and freeze in scenarios. They share a similar temperature to the Standard Model plasma and would become non-relativistic when the temperature falls below their mass $m_4$.
In this work, we will focus on the mass range $m_4>{\rm MeV}$ and explore the decay effects on BBN at temperatures below MeV.
Therefore, for the entire parameter space of interest to this study, the sterile neutrinos are produced relativistically but decay after they have become non-relativistic. There is always sufficient expansion of the universe for the transition to occur. 

Thanks to this separation of scales, we can simply account for the heavy neutrino decay by multiplying an exponential decay factor to its number density.
The number density at late time is given by
\begin{equation}\label{eq:DecayFactor}
n_{\nu_4}(t) = n_{\nu_4}(t_i) \left(\frac{a(t_i)}{a(t)}\right)^3 \exp\left(- \frac{\Gamma_{\nu_4}^{\rm tot}}{2H}\right) \ ,
\end{equation}
where $\Gamma_{\rm tot}$ is the total decay rate of heavy neutrino $\nu_4$ and $t_i$ is a time chosen to be well after the production and well before the decay.

\begin{figure}
    \centering
    \includegraphics[width = 1.0\linewidth]{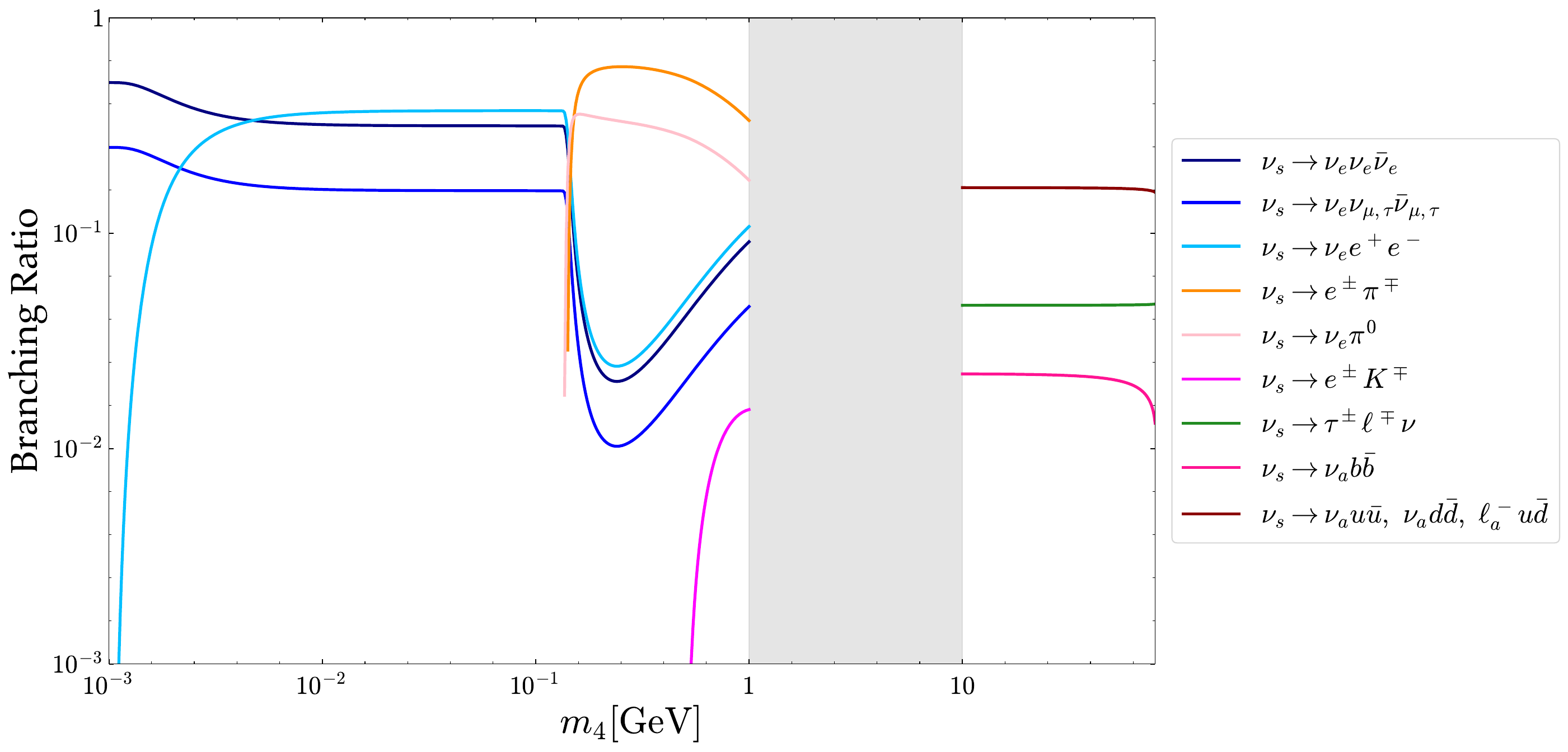}
    \caption{Branching ratio of heavy neutrino as a function of its mass used in our analysis (which is independent of the mixing parameter $|U_{a4}|^2$). For clarity, we show the decay modes most relevant for our BBN analysis. We assume the heavy neutrino to be Majorana and for each decay channel the charge-conjugation (if exist) is already included.}
    \label{fig: BR}
\end{figure}

The heavy neutrino can decay in leptonic and semi-leptonic modes. Important decay branching ratios for our analysis is shown in Fig.~\ref{fig: BR}.
Neglecting the final state particle masses, the partial decay width in the rest frame of a light $\nu_4$ takes the generic form
\begin{equation}
    \Gamma_{\nu_4\to X} = \tilde c_{\nu_4\to X} \dfrac{G_F^2 m_4^5}{\pi^3} |U_{a4}|^2 \ .
\label{eq: decay width}
\end{equation}
From Eq.~\eqref{eq:T*}, we find that $T_*$ is always much larger than the BBN temperatures, for the mass range $m_4>{\rm MeV}$ we will consider. 
Therefore, the active-sterile mixing angle governing the decay takes the vacuum value to a very good approximation, i.e., $\sin^2(2\theta_{\rm eff}) \simeq \sin^2(2\theta) \simeq 4 |U_{a4}|^2$.

The decay coefficients for leptonic decay modes are
\begin{equation}
\label{eq:LeptonicDecayCoeff}
\begin{split}
\tilde c_{\nu_4 \to \nu_a \nu_a \bar{\nu}_a} &= \frac{1}{384} \ , \quad
\tilde c_{\nu_4 \to \nu_a \ell^+_a \ell^-_a} = \frac{5}{1536} \ , \\
\tilde c_{\nu_4 \to \nu_a \nu_b \bar{\nu}_b} &= \frac{1}{768} \ , \quad
\tilde c_{\nu_4 \to \nu_a \ell^+_b \ell^-_b} = \frac{1}{1536} \ , \quad
\tilde c_{\nu_4 \to \nu_b \ell^+_b \ell^-_a} = \frac{1}{192} \ , \\
\end{split}
\end{equation}
Here the flavor indices are not summed over, we assume that $\nu_4$ mixes with $\nu_a$, and $b\neq a$.
For simplicity, we make the approximation that $\sin^2\theta_W \simeq 1/4$.
It is worth noting that when all final state neutrino(s) and charged leptons have the same flavor, there are interference effects between Feynman diagram contributions.
Kinematically, for $\nu_4$ mass below the pion mass threshold, it can only decay into light neutrinos and $e^\pm$.

For $\nu_4$ mass above the pion or kaon mass thresholds but below GeV scale, the dominant heavy neutrino decay modes are into a  lepton and a meson $\mathfrak{m} = \pi, K$. The corresponding decay coefficients are~\cite{Bondarenko:2018ptm},
\begin{equation}\label{eq:nu4tomeson}
\begin{split}
\tilde c_{\nu_4 \to \nu_a \pi^0} &= \frac{\pi^2 f_\pi^2}{32 m_4^2} \left(1- \frac{m_\pi^2}{m_4^2} \right)^2 \ , \\
\tilde c_{\nu_4 \to \ell^-_a \pi^+} &= \frac{\pi^2 f_\pi^2  |V_{ud}^{\rm CKM}|^2}{16 m_4^2} \left[\left(1- \frac{m_\ell^2}{m_4^2} \right)^2- \frac{m_\pi^2}{m_4^2}\left(1+ \frac{m_\ell^2}{m_4^2} \right)^2 \right]
\sqrt{\lambda\left(1, \frac{m_\ell^2}{m_4^2}, \frac{m_\pi^2}{m_4^2}\right)} \ , \\
\tilde c_{\nu_4 \to \ell^-_a K^+} &= \frac{\pi^2 f_K^2  |V_{us}^{\rm CKM}|^2}{16 m_4^2} \left[\left(1- \frac{m_\ell^2}{m_4^2} \right)^2- \frac{m_K^2}{m_4^2}\left(1+ \frac{m_\ell^2}{m_4^2} \right)^2 \right]
\sqrt{\lambda\left(1, \frac{m_\ell^2}{m_4^2}, \frac{m_K^2}{m_4^2}\right)} \ , 
\end{split}
\end{equation}
where $\lambda(\alpha, \beta, \gamma) = \alpha^2+\beta^2+\gamma^2 -2\alpha\beta - 2 \alpha\gamma-2\beta\gamma$ and $V_{ij}^{\rm CKM}$ is CKM matrix element.

For the $\nu_4$ mass well above the GeV scale, one must consider inclusive quark final states.
In the heavy $W,Z$ limit, the decay coefficients for three-body semi-leptonic decays are
\begin{equation}\label{eq:nu4toquark}
\begin{split}
\tilde c_{\nu_4 \to \nu_a u_i \bar{u}_i} &= \frac{5}{2304} \ , \quad
\tilde c_{\nu_4 \to \nu_a d_i \bar{d}_i} = \frac{13}{4608} \ , \quad
\tilde c_{\nu_4 \to u_i \bar d _j \ell^-_a} = \frac{|V_{ij}^{\rm CKM}|^2}{64} \ . \\
\end{split}
\end{equation}
For very heavy $\nu_4$ with mass close to the weak boson masses, one cannot work within the Fermi theory and the corresponding decay rates are presented in appendix~\ref{app:B}.

With the knowledge of the heavy neutrino decay products and branching ratios, we proceed to explore their impact on BBN.
As a brief outline, our discussion in sections~\ref{sec:chargedmeson} -- \ref{sec:nuinj} will fucus on heavy neutrinos lighter than a few GeV scale, where exclusive decays into light mesons dominate. 
We adopt a semi-analytic approach to provide an anatomy of their effects on BBN from various reactions at the nucleon and nucleus levels.
The treatment for the large mass region ($m_4\gg{\rm GeV}$) will be discussed in section~\ref{sec:largemass}.

\subsection{Charged pion and kaon injection}\label{sec:chargedmeson}

Hadronic injection is well known to have the strongest effect on BBN.
We first consider the mass of heavy neutrino to be above the QCD scale but lighter than a few GeV, where the dominant decay modes of heavy neutrino are into a lepton plus a pion or kaon.
The corresponding charged pion and kaon injections from heavy neutrino decay can cause proton-to-neutron ($pn$) conversion as well as hadronic dissociation of nuclei through strong interactions, which are the two most important effects strongly constrained by BBN.

Thanks to the fact that the charged pion/kaon lifetimes are much shorter than duration of BBN, our analysis can be simplified by only considering their instantaneous injection from heavy neutrino decay.
In other words, the charged mesons that cause $pn$ conversion or nuclear dissociation at time $t$ during BBN are exclusively from the decay of heavy neutrino at almost the same time $t$.

\subsubsection{Charged meson injection rates}

For heavy neutrino mass below a few GeV scale, we can use the two-body partial decay rate Eq.~\eqref{eq:nu4tomeson} to derive the charged kaon injection rate,
\begin{equation}\label{eq:3.6}
\mathcal{R}_{\nu_4\to K^\pm}(t) = n_{\nu_4}(t) \Gamma_{\nu_4\to \ell_a^\mp K^\pm} \ ,
\end{equation}
where the factor $n_{\nu_4}$ is obtained using Eq.~\eqref{eq:DecayFactor}.
For charged pions, we sum up the contribution from direct decay of heavy neutrino and the cascade decay via charged kaons,
\begin{equation}\label{eq:3.7}
\mathcal{R}_{\nu_4\to \pi^\pm}(t) = 
n_{\nu_4}(t) \Gamma_{\nu_4\to \ell_a^\mp \pi^\pm} + 
n_{\nu_4}(t) \Gamma_{\nu_4\to \ell_a^\mp K^\pm} \text{Br}(K^\pm \to \pi^\pm) \ .
\end{equation}
The produced $K^\pm$ and $\pi^\pm$ typically slow down and thermalize in the early universe plasma, before decaying away or interacting with other hadrons~\cite{Pospelov:2010cw}. As a result, we only need to obtain the overall rate of kaon and pion injection instead of their energy spectra.
We do not consider $pn$ conversion triggered by neutral pions because they decay at much shorter lifetime. The effects of photons resulting from $\pi^0$ will be addressed later on.
We also do not consider neutral kaons whose production from heavy neutrino decay is GIM suppressed.

\subsubsection{Proton-neutron conversion with charged mesons}
\label{subsec: Nucleon Conversion}

The first hadronic effect of heavy neutrino decay in our analysis is the $pn$ conversion induced by charged mesons. Although the kaons are produced with a CKM suppression factor $|V_{us}|^2$ (see also Fig.~\ref{fig: BR}), their interaction cross sections with nucleons are much larger than those of pions. Thus we start with kaons.
The kaon-nucleon reaction occurs through the hyperon intermediate states, resulting in multiple pions in the final state~\cite{Pospelov:2010cw}. 
It is sufficient to consider the effects from $K^-$.
The $K^+p$ interaction is suppressed due to the Coulomb screening whereas the $K^+n$ interaction (through intermediate $\Sigma^+ \pi^0, \Sigma^0 \pi^+$ states) is forbidden in the limit of isospin conservation.

The Boltzmann equation for proton and neutron due to $K^-$ injection are
\begin{equation}\label{eq:BoltzmannPNkaon}
\begin{split}
    \frac{dn_{p}}{dt} \simeq \left(\dfrac{dn_{p}}{dt} \right)_{\rm SBBN} &
    + \mathcal{R}_{\nu_4\to K^-}(t) \times
    \frac{ n_{n}(t) \langle \sigma v \rangle_{K^- n \to p X}
    }{
    \tau_{K^-}^{-1}  + \Gamma_{K^-}^\text{conv} + \Gamma_{K^-}^\text{diss} + \Gamma_{K^-}^\text{anni}
    } \\
    &- \mathcal{R}_{\nu_4\to K^-}(t) \times
    \frac{ 
    n_{p}(t) \langle \sigma v \rangle_{K^- p \to n X}
    }{
    \tau_{K^-}^{-1}  + \Gamma_{K^-}^\text{conv} + \Gamma_{K^-}^\text{diss} + \Gamma_{K^-}^\text{anni}
    } + \hdots \ , \\
    \frac{dn_{n}}{dt} \simeq
    \left(\frac{dn_{n}}{dt}\right)_{\rm SBBN} &
    - \mathcal{R}_{\nu_4\to K^-}(t) \times
    \frac{ n_{n}(t) \langle \sigma v \rangle_{K^- n \to p X}
    }{
    \tau_{K^-}^{-1}  + \Gamma_{K^-}^\text{conv} + \Gamma_{K^-}^\text{diss} + \Gamma_{K^-}^\text{anni}
    } \\
    &+ \mathcal{R}_{\nu_4\to K^-}(t) \times
    \frac{ 
    n_{p}(t) \langle \sigma v \rangle_{K^- p \to n X}
    }{
    \tau_{K^-}^{-1} + \Gamma_{K^-}^\text{conv} + \Gamma_{K^-}^\text{diss} + \Gamma_{K^-}^\text{anni}
    } + \hdots \ ,
\end{split}
\end{equation}
where $X$ indicates various multi-pion final states.
On the right-hand side of both equations, the first term denotes all the processes in the standard BBN (SBBN). The new effects due to heavy neutrino to $K^-$ decay are encoded by the second and third terms.
The kaon injection rate $\mathcal{R}_{\nu_4\to K^-}(t)$ is defined in Eqs.~\eqref{eq:3.6}. 
The $\dots$ represents the other reactions of kaon and other decay products of the heavy neutrino that contribute to the rate equations (see the following subsections).

The fractional factors multiplying  with $\mathcal{R}$ show the competition among various $K^-$ reaction channels.
In the numerators, the conversion cross sections for thermalized kaons are~\cite{Pospelov:2010cw}
\begin{equation}
        \langle \sigma v \rangle_{K^- p \to n X} \simeq 32 \,\text{mb} \ , \quad
        \langle \sigma v \rangle_{K^- n \to p X} \simeq 13 \,\text{mb} \ .
\end{equation}
We use the SBBN solutions for the proton and neutron number densities, $n_{p,n}(t)$, which acknowledges the fact that new physics is not permitted to cause significant departure from SBBN predictions.
In the denominators, $\tau_{K^-} = 1.24\times 10^{-8}\,{\rm sec}$ is the charged kaon lifetime in the rest frame.
The kaon induced $pn$ conversion rate is given by the sum of numerators
\begin{equation}
\Gamma_{K^-}^\text{conv} = n_{n}(t) \langle \sigma v \rangle_{K^- n \to p X} + n_{p}(t) \langle \sigma v \rangle_{K^- p \to n X} \ .
\end{equation}
The $K^-$ induced nuclei dissociation rate $\Gamma_{K^-}^\text{diss}$ will be presented in the upcoming section~\ref{subsec: Helium-4 Dissociation}.

Other than the decay and reactions with nucleon and nuclei, the last term in the denominator of Eq.~\eqref{eq:BoltzmannPNkaon}, $\Gamma_{K^-}^\text{anni}$, represents the self-annihilation channels  that can also destroy the kaons. Because the thermal population of kaon is highly Boltzmann suppressed at BBN times, both $K^\pm$ participating the self-annihilation are from the heavy neutrino decay. 
The two kaons mainly annihilate into pions and the cross section is roughly $\langle \sigma v \rangle_{\rm anni} \sim m_{K^-}^{-2}$~\cite{Akita:2024ork}.
The kaon self-annihilation rate can be estimated as the following,
$\Gamma_{K^-}^\text{anni} = n_{K^-} \langle \sigma v \rangle_{\rm anni}$, where the instantaneous kaon number density is bounded from above $n_{K^-} \sim \mathcal{R}_{\nu_4\to K}/(\tau_{K^-}^{-1}  + \Gamma_{K^-}^\text{conv} + \Gamma_{K^-}^\text{diss} + \Gamma_{K^-}^\text{anni}) < \tau_{K^-}\mathcal{R}_{\nu_4\to K}  = n_{\nu_4} {\rm Br}_{\nu_4\to K} (\tau_{K^-}/\tau_{\nu_4})$. 
Using Fig.~\ref{fig: BR}, we find the branching ratio to kaon is smaller than $10^{-2}$.
Moreover, because the population of heavy neutrino cannot exceed the thermal value, i.e., $n_{\nu_4} \lesssim T^3$, $\Gamma_{K^-}^\text{anni}$ is further bounded from above as
\begin{equation}\label{eq:uppergoundGammaSelf}
\Gamma_{K^-}^\text{anni} < 10^{-8} \frac{T^3}{m_{K^-}^2} \left(\frac{0.02\,{\rm sec}}{\tau_{\nu_4}}\right) \simeq 10^{-17} {\rm GeV} \left( \frac{T}{\rm MeV} \right)^3 \left(\frac{0.02\,{\rm sec}}{\tau_{\nu_4}}\right) \ .
\end{equation}
The reference point for $\tau_{\nu_4}$ is the shortest lifetime of heavy neutrino that can be constrained in this work. Correspondingly, the highest possible value of $\Gamma_{K^-}^\text{anni}$ from Eq.~\eqref{eq:uppergoundGammaSelf} is below the kaon decay width. Based on this estimate, 
we conclude that self-annihilation effect of kaons can be safely neglected for the heavy neutrino model being explored here.

Similar to the discussion for kaons, the Boltzmann equations for proton and neutron due to charged pion induced $pn$ conversion effect read
\begin{equation}\label{eq:BoltzmannPNpion}
\begin{split}
    \frac{dn_{p}}{dt} \simeq \left(\dfrac{dn_{p}}{dt} \right)_{\rm SBBN} &+ \mathcal{R}_{\nu_4\to \pi^+}(t) \times
    \frac{ n_{n}(t) \langle \sigma v \rangle_{\pi^+ n \to p \pi^0}
    }{
    \tau_{\pi^+}^{-1}  + \Gamma_{\pi^+}^\text{conv} + \Gamma_{\pi^+}^\text{anni}
    } \\
    & - \mathcal{R}_{\nu_4\to \pi^-}(t) \times
    \frac{ 
    n_{p}(t) \left( \langle \sigma v \rangle_{\pi^- p \to n \gamma} + \langle \sigma v \rangle_{\pi^- p \to n \pi^0} \rule{0mm}{4mm}\right) 
    }{
    \tau_{\pi^-}^{-1}  + \Gamma_{\pi^-}^\text{conv} + \Gamma_{\pi^-}^\text{diss} + \Gamma_{\pi^-}^\text{anni}
    } +\cdots \ , \\
    \frac{dn_{n}}{dt} = 
    \left(\frac{dn_{n}}{dt}\right)_{\rm SBBN} &- 
    \mathcal{R}_{\nu_4\to \pi^+}(t) \times
    \frac{ n_{n}(t) \langle \sigma v \rangle_{\pi^+ n \to p \pi^0}
    }{
    \tau_{\pi^+}^{-1}   + \Gamma_{\pi^+}^\text{conv} + \Gamma_{\pi^+}^\text{anni}
    } \\
    & + \mathcal{R}_{\nu_4\to \pi^-}(t) \times
    \frac{ 
    n_{p}(t) \left( \langle \sigma v \rangle_{\pi^- p \to n \gamma} + \langle \sigma v \rangle_{\pi^- p \to n \pi^0} \rule{0mm}{4mm}\right) 
    }{
    \tau_{\pi^-}^{-1} + \Gamma_{\pi^-}^\text{conv} + \Gamma_{\pi^-}^\text{diss} + \Gamma_{\pi^-}^\text{anni}
    }+\cdots \ ,
\end{split}
\end{equation}
where the charged pion injection rate $\mathcal{R}_{\nu_4\to \pi^\pm}(t)$ is given by Eqs.~\eqref{eq:3.7}.
The $pn$ conversion cross sections triggered by thermalized pions are~\cite{Kohri:2001jx, Kawasaki:2004qu, Pospelov:2010cw}
\begin{equation}
         \langle \sigma v \rangle_{\pi^- p \to n \gamma} \simeq 0.57 \,\text{mb} \ , \quad
        \langle \sigma v \rangle_{\pi^- p \to n \pi^0} \simeq 0.88 \,\text{mb} \ , \quad
        \langle \sigma v \rangle_{\pi^+ n \to p \pi^0} \simeq 1.7 \,\text{mb} \ ,
\end{equation}
and for the nucleon number densities $n_{p,n}(t)$ we again use their SBBN values.
In the denominator, $\tau_{\pi^\pm} = 2.6\times 10^{-8}\,{\rm sec}$ is the charged pion lifetime. The conversion rates are 
\begin{equation}
\begin{split}
\Gamma_{\pi^+}^\text{conv} &= n_{n}(t) \langle \sigma v \rangle_{\pi^+ n \to p \pi^0} \ , \\
\Gamma_{\pi^-}^\text{conv} &= n_{p}(t) \left( \langle \sigma v \rangle_{\pi^- p \to n \gamma} + \langle \sigma v \rangle_{\pi^- p \to n \pi^0} \right) \ .
\end{split}
\end{equation}
The pion induced dissociation rates of nuclei $\Gamma_{\pi^-}^\text{diss}$ will be presented in section~\ref{subsec: Helium-4 Dissociation}.

For the self-annihilation of $\pi^\pm$, the annihilation cross section into $2\pi^0$ is phase space suppressed whereas annihilation into two photons is suppressed by the small electromagnetic coupling $\alpha^2$. Similar to the argument for the kaon case, we find $\Gamma_{\pi^\pm}^\text{anni}$ is numerically also negligible compared to the other reaction rates in the denominator. 

\subsubsection{Hadrodissociation of ${}^4{\rm He}$ with charged mesons}
\label{subsec: Helium-4 Dissociation}

The other important effect of heavy neutrino decay into charged mesons is their hadrodissociation. In this work, we focus on Helium-4 (${}^4{\rm He}$), which has the largest abundance other than protons, the largest hadrodissociation cross section, and rich daughter particles after the dissociation.
The Boltzmann equation for ${}^4{\rm He}$ due to $K^-, \pi^-$ dissociation is
\begin{equation}\label{eq:BoltzmannHekaon}
\begin{split}
    \frac{dn_{\rm {}^4He}}{dt} \simeq \left(\dfrac{dn_{\rm {}^4He}}{dt} \right)_{\rm SBBN} &-
    \mathcal{R}_{\nu_4\to K^-}(t) \times
    \frac{ n_{\rm {}^4He}(t) \sum_N \langle \sigma v \rangle_{K^- {\rm {}^4He} \to N X}
    }{
    \Gamma_{K^-}  + \Gamma_{K^-}^\text{conv} + \Gamma_{K^-}^\text{diss} + \Gamma_{K^-}^\text{anni}
    } \\
    &-\mathcal{R}_{\nu_4\to \pi^-}(t) \times
    \frac{ n_{\rm {}^4He}(t) \sum_N \langle \sigma v \rangle_{\pi^- {\rm {}^4He} \to N X}
    }{
    \Gamma_{\pi^-}  + \Gamma_{\pi^-}^\text{conv} + \Gamma_{\pi^-}^\text{diss} + \Gamma_{\pi^-}^\text{anni}
    }
    + \hdots \ ,
\end{split}
\end{equation}
where $N$ labels various nucleus/nucleon final states. The cross sections are~\cite{PhysRevD.1.1267, Pospelov:2010cw, Daum:1995au}.
\begin{equation}
\begin{split}
&\langle \sigma v \rangle_{K^- {\rm {}^4He} \to N X} \simeq 60 \xi_N \ , \quad
\xi_{^3 {\rm He}} \simeq \xi_{\rm T} \simeq 0.13 \ , \quad 
\xi_{\rm D} \simeq 0.34 \ , \quad 
\xi_n \simeq 1.4 \ , \quad 
\xi_p \simeq 1.12 \ , \\
&\langle \sigma v \rangle_{\pi^- {\rm {}^4He} \to {\rm T} n} \simeq 1.1 \,\text{mb} \ , \quad
        \langle \sigma v \rangle_{\pi^- {\rm {}^4He} \to {\rm D} 2n} \simeq 4.1 \,\text{mb} \ , \quad
        \langle \sigma v \rangle_{\pi^+ {\rm {}^4He} \to p 3n} \simeq 1.3 \,\text{mb} \ .
\end{split}
\end{equation}
The above Born-level cross sections are further multiplied by an $S$-wave Sommerfeld factor due to the opposite sign of $K^-$ and ${}^4{\rm He}$ nucleus charges~\cite{Arkani-Hamed:2008hhe, Feng:2009mn, Reno:1987qw, Pospelov:2010cw}
\begin{equation}
    S = \frac{2\pi Z \alpha/v}{1-e^{-2 \pi Z\alpha/v}} \ ,
\end{equation}
where $v = \sqrt{2T/\mu}$ is the thermally-averaged relative velocity between $K^-$ and ${}^4{\rm He}$ and $\mu$ is the reduced mass of scattering. 
The factor of $Z=2$ is the electric charge carried by ${}^4{\rm He}$, in unit of $e$.

Like before, we use the SBBN values for the number density $n_{\rm {}^4He}(t)$.
We only consider dissociation with $K^-, \pi^-$ particles because the cross sections for $K^+,\pi^+$ are suppressed due to Coulomb screening.
The rates for regular decay, $pn$ conversion, and self-annihilation for kaon and pion have been discussed in the previous subsection. Here, we present their dissociation rates of ${}^4{\rm He}$ used in our analysis
\begin{equation}
\begin{split}
    \Gamma_{K^-}^\text{diss} &= n_{\rm {}^4He}(t) \sum_{N = {}^3{\rm He}, {\rm T}, {\rm D}, n, p} \langle \sigma v \rangle_{K^- {\rm {}^4He} \to N X} \ ,\\
    \Gamma_{\pi^-}^\text{diss} &= n_{\rm {}^4He}(t) \sum_{N = {\rm T}, {\rm D}, p} \langle \sigma v \rangle_{\pi^- {\rm {}^4He} \to N X} \ .
\end{split}
\end{equation}

\subsection{Photon injection}\label{sec:photoninj}

Electromagnetic injection is another potentially important effect of new physics on BBN. 
The photons could potentially be absorbed by the primordial elements and dissociate them into nucleons or lighter nuclei. During the BBN time, photons are tightly coupled to the background thermal plasma through Compton scatterings and the Breit–Wheeler processes with a large optical depth. 
As a result, the non-thermal photon lifetime is very short.
Similar to the approximation made in the charged pion case, for the photon injection rate we only consider the instantaneous decay of the heavy neutrino.

\subsubsection{Photon injection energy spectrum}\label{sec:3.2.1}

The photon triggered nucleus dissociation cross sections are energy dependent thus we will need to find out the energy spectrum of photons from heavy neutrino decay. For heavy neutrino mass below the GeV scale, the dominant cascade process is $\nu_4 \to \nu_a \pi^0$, followed by $\pi^0 \to\gamma\gamma$. In the first step, because the heavy neutrino is already non-relativistic when decaying, the resulting $\pi^0$ has approximately fixed energy $E_{\pi} = (m_4^2 + m_{\pi}^2)/(2m_4)$. In the boosted-pion reference frame, the resulting photon energy spectrum is
\begin{equation}\label{eq:GammaSpectrum1}
\frac{d\mathcal{R}_{\nu_4\to \gamma}}{d E_\gamma} = n_{\nu_4}(t) \Gamma_{\nu_4 \to \nu_a \pi^0} \frac{2}{E_\pi \beta_\pi} \Theta\left(E_\gamma - (E_\gamma)_{\rm min}\rule{0mm}{4mm}\right)\Theta\left((E_\gamma)_{\rm max}-E_\gamma\rule{0mm}{4mm}\right)
\ ,
\end{equation}
where $\beta_\pi = (m_4^2 - m_{\pi}^2)/(m_4^2 + m_{\pi}^2)$, $\Theta$ is the unit-step function, and we observe that the photons have a flat energy distribution with 
\begin{equation}
(E_\gamma)^{\rm max}_{\rm min} = \frac{E_\pi}{2} (1\pm \beta_\pi) \ .
\end{equation}

\subsubsection{Photodissociation of nuclei}
\label{subsec: Deuterium Dissociation}

Here, we present the set of Boltzmann equations that include the non-thermal photodissociation effects.
For simplicity of discussion, we only add new terms related to deuterium destruction. 
Our numerical analysis goes beyond this  and includes nine light nuclei dissociation channels up to ${}^4{\rm He}$~\cite{Cyburt:2002uv}. 
The corresponding cross sections are listed in the appendix~\ref{app:C}. 

The deuterium dissociation process is $d + \gamma \to p + n$, which is inverse to the first step of fusion processes of BBN. It depletes deuterium in the universe and contributes to the populations of proton and neutron.
In this case, the modified Boltzmann equations are
\begin{equation} \label{eq: Boltzmann for deuterium}
\begin{split}
    \frac{d n_d}{dt}
    &= \left( \frac{d n_d}{dt} \right)_{\rm SBBN} - \int_Q^\infty d E_\gamma
    \frac{d \mathcal{R}_{\nu_4\to\gamma}}{d E_\gamma}
    \frac{n_d \langle \sigma v \rangle_{d\gamma \to np}}{\Gamma^\gamma_{\rm tot}} + \cdots \ , \\
    \frac{d n_{p,n}}{dt}
    &= \left( \frac{d n_{p,n}}{dt} \right)_{\rm SBBN} + \int_Q^\infty d E_\gamma
    \frac{d \mathcal{R}_{\nu_4\to\gamma}}{d E_\gamma}
    \frac{n_d \langle \sigma v \rangle_{d\gamma \to np}}{\Gamma^\gamma_{\rm tot}} + \cdots \ ,
\end{split}    
\end{equation}
where the non-thermal photon energy spectrum $d \mathcal{R}_{\nu_4\to\gamma}/dE_\gamma$ is given above in Eq.~\eqref{eq:GammaSpectrum1}, and 
\begin{equation}\label{eq:GammaTotPhoton} 
\Gamma^\gamma_{\rm tot} = n_d (\sigma v)_{d\gamma \to np} + n_\gamma (\sigma v)_{\gamma\gamma \to e^+ e^-} + n_e (\sigma v)_{e^\pm \gamma \to e^\pm \gamma} + \cdots \ .
\end{equation}
On the right-hand side, the deuterium dissociation cross section is
\begin{equation}
\sigma_{d \gamma \to n p} (E_\gamma) = 18.75 \textrm{mb} 
    \left[ 
    \left(
    \frac{\left| Q \right| ( E_\gamma - \left| Q \right| )}{E_\gamma}
    \right)^3
    +0.007947
    \left(
    \frac{\left| Q \right| ( E_\gamma - \left| Q \right| )}{E_\gamma}
    \right)^2
    \frac{
    \left(
    \sqrt{\left| Q \right|} - \sqrt{0.037}
    \right)^2
    }{
    E_\gamma - ( \left| Q \right| - 0.037)
    }
    \right] \ ,
\end{equation}
where $Q=2.2\,$MeV is the deuterium binding energy, and we use the standard BBN values for $n_d$, $n_\gamma$, $n_e$ to evaluate the non-thermal photon reaction rates.
$\gamma\gamma \to e^+ e^-$ and $e^\pm \gamma \to e^\pm \gamma$ stand for the regular QED processes where the energetic photon from heavy neutrino decay scatters with a background electron (Compton scattering~\cite{Peskin:1995ev}) and photon (Breit–Wheeler scattering~\cite{Ribeyre:2015fta}) from the thermal plasma, respectively. 
The dots in Eqs.~\eqref{eq: Boltzmann for deuterium} and \eqref{eq:GammaTotPhoton}  represents other nuclei photodissociation rates taken into account by our analysis.

\subsection{Neutrino injection}\label{sec:nuinj}

We also consider active neutrinos from the heavy neutral lepton decay. 
Like charged pions, neutrinos can also cause the $pn$ conversion.
Due to the weakly interacting nature, the conversion effect from neutrinos is weaker than pions if both are present from the decay of heavy neutrino. However, if the heavy neutrino mass lies below the pion decay threshold, it can only decay into light neutrinos and $e^\pm$. This is the mass region where neutrinos could play a leading role in modifying BBN predictions.

\subsubsection{Neutrino injection energy spectrum}

Different from charged pions and photons, neutrino does not decay and interacts feebly. This leads to an accumulation effect in the number of neutrinos. At any given time, the neutrinos participating in the $pn$ conversion could be produced from heavy neutrino decays in the past, roughly, one mean free path back in time.

We following the general approach described in~\cite{Nemevsek:2023yjl} to derive the phase space distribution function of active neutrinos from heavy neutrino decay,
\begin{equation}\label{eq:NuPSD}
f_{\nu_e} (x, T) =  \frac{2\pi^2}{x^2} \int_T^{T_{\rm max}(x)} \frac{dT'}{T'} \frac{n_{\nu_4}(T')}{m_4 H(T') T'^2} \sum_i  \Gamma_{\nu_4\to \nu}^i g_i\left(\frac{T'}{m_4}x\right) \ ,
\end{equation}
where $x=E_\nu/T$, the sum over $i$ goes over the leptonic decay channels in Eq.~\eqref{eq:LeptonicDecayCoeff} that can produce $\nu_e$ in the final state. $\Gamma_{\nu_4\to \nu}^i$ is the partial decay rate for one of the decay modes there, and $g_i$ is corresponding dimensionless neutrino spectral function per heavy neutrino decay,
\begin{equation}
g_i(\omega) = \frac{n}{\Gamma_{\nu_4\to \nu}^i} \frac{d\Gamma_{\nu_4\to \nu}^i}{d\omega} \ , \quad \int d\omega g(\omega) = n \ ,
\end{equation}
where $\omega = E_\nu/m_4$ and $n$ is the number of $\nu_e$ produced in the decay mode. For the $\nu_4\to \nu_a\nu_a \bar\nu_a$ decay channel, $n=2$. For the other channels $n=1$. For weak-boson mediated decay modes in Eq.~\eqref{eq:LeptonicDecayCoeff}, the $g_i$ functions take a universal form,
\begin{equation}
g_i(\omega) = 16n \omega^2 (3-4 \omega) \ .
\end{equation}

The temperature upper limit $T_{\rm max}$ of the integral of Eq.~\eqref{eq:NuPSD} is 
\begin{equation}
T_{\rm max}(x) \simeq \frac{1.3}{\left(G_F^2 M_{\rm P} x\rule{0mm}{3mm}\right)^{1/3}} \ .
\end{equation}
It is set by $\Gamma_a = H$ where $\Gamma_a$ is the reaction rate of an active neutrino with energy $E_\nu=xT$. 
By imposing this upper limit, our analysis only consider the fraction of active neutrinos produced from $\nu_4$ decay that will free stream in the universe, except for occasionally striking on the remaining nucleon in the universe to trigger the $pn$ conversion.
If the heavy neutrino is sufficiently heavy and long lived, it can produce more energetic active neutrinos which are not decoupled from the thermal plasma. The corresponding dynamics is complex and need to be understood numerically~\cite{Ruchayskiy:2012si,Sabti:2020yrt,Boyarsky:2020dzc,Boyarsky:2021yoh}. In order to proceed with our semi-analytic discussion, we neglect the very energetic neutrino which leads to a more conservative (weaker) constraint. 
In practice, we find the neutrino injection effect only stands out for heavy neutrino below $\sim 100$ MeV. In the same mass region, the constraint on smaller mixings is overshadowed by a complementary effect from temporary matter domination discussed in section~\ref{subsec: Temporary matter domination}.
See Fig.~\ref{fig:MainPlotE2} and discussions therein.

The heavy neutrino considered in this work experiences no CP violation in its tree-level production and decay. As a result, the resulting neutrino and anti-neutrino distributions are identical
\begin{equation}
f_{\bar\nu_e} = f_{\nu_e} \ .
\end{equation}

\subsubsection{Proton-neutron conversion with \texorpdfstring{$\nu_e$}.}
\label{subsec: Proton-neutron conversion triggered by neutrino injection}

Like charged pions, neutrino injection can also trigger the $pn$ conversion via weak interactions. 
Due to the large mass of muon and tau leptons, we focus on the $\nu_e$ neutrinos for the $pn$ conversion process.
With the neutrino PSD function derived in Eq.~\eqref{eq:NuPSD}, the Boltzmann equations for proton and neutron are as follows
\begin{equation}
\begin{split}
    \dfrac{dn_p}{dt} 
    &= \left( \dfrac{dn_p}{dt} \right)_{\rm SBBN} - n_p\int
    \dfrac{d^3 p_{\bar{\nu}_e}}{(2 \pi)^3}
    f_{\bar{\nu}_e} \sigma_{p \bar{\nu} \to n e^+} 
    + n_n\int
    \dfrac{d^3 p_{\nu_e}}{(2 \pi)^3}
    f_{\nu_e} \sigma_{n \nu \to p e^-} \ , \\
    \dfrac{dn_n}{dt} 
    &= \left( \dfrac{dn_n}{dt} \right)_{\rm SBBN} + n_p\int
    \dfrac{d^3 p_{\bar{\nu}_e}}{(2 \pi)^3}
    f_{\bar{\nu}_e} \sigma_{p \bar{\nu} \to n e^+} 
    - n_n\int
    \dfrac{d^3 p_{\nu_e}}{(2 \pi)^3}
    f_{\nu_e} \sigma_{n \nu \to p e^-} \ .
\label{eq: Boltzmann for n to p in small mass}
\end{split}
\end{equation}
With the assumption of $E_\nu \gg m_e$ and $m_n \simeq m_p \equiv m_N$, the quasi-elastic nucleon-conversion cross section take the approximate form (in the rest frame of the initial-state nucleon)
\begin{equation}
\begin{aligned}
    (\sigma v)_{p \bar{\nu}_e \to n e^+} &\simeq
    \frac{4 G_F^2E_\nu^2 m_N \left[ 16(g_A^2 - g_A +1) E_\nu^2 + 12 (2 g_A^2 - g_A + 1) m_N E_\nu + 3(3 g_A^2+1) m_N^2 \right] }{3\pi(m_N+2E_\nu)^3} \ , \\
    (\sigma v)_{n \nu_e \to p e^-} &\simeq 
    \frac{4 G_F^2 E_\nu^2 m_N  \left[ 16(g_A^2 + g_A +1) E_\nu^2 + 12 (2 g_A^2 + g_A + 1) m_N E_\nu + 3(3 g_A^2+1) m_N^2 \right]}{3\pi(m_N+2E_\nu)^3} \ ,
\end{aligned}
\end{equation}
where $g_A=1.27$ is the nucleon axial-vector current coupling constant.
In our numerical simulation, we keep the electron mass and the proton-neutron mass difference, which set the neutrino energy threshold for the $p \bar{\nu}_e \to n e^+$ reaction to occur, 
\begin{equation}
(E_\nu)_{\rm min}\simeq\frac{(m_n-m_p)^2-m_e^2}{2(m_n-m_p)} \ .
\end{equation}
As a useful remark on the details, even in the case where the heavy neutrino only mixes with $\nu_\mu$ or $\nu_\tau$ and it is lighter than the corresponding charged lepton, weak interaction still allows it decay into $\nu_e$ via the $Z$-boson exchange.

\subsection{The large mass region ($m_4>10\,$GeV)}\label{sec:largemass}

So far, our discussions of the heavy neutrino decay effects on BBN have been based on a simplified approach where the injected particles are either short-lived ($K^\pm, \pi^\pm, \pi^0$), strongly interacting with the thermal plasma ($\gamma$), or free streaming ($\nu$'s). These features allow us to analytically account for the non-thermal effects due to heavy neutrino decay as new source terms in the Boltzmann equations for proton, neutron, and nuclei.
Such an approach is most powerful for exploring heavy neutrino with mass below a few GeV, where its leptonic and hadronic decay channels are calculated exclusively.

Next, we move on to the BBN constraint for heavy neutrino with mass above 10 GeV. In this mass range, its semi-leptonic decays must be calculated inclusively with quark and anti-quark final states at short distances. The subsequent hadronization occurs non-perturbatively and can produce a large sample of baryons and mesons. These non-thermal, energetic particles can further interact with the thermal plasma to produce secondary particles, whereas the antibaryons can also deplete baryons from the nuclei. 
Such complexity is difficult to handle using our semi-analytic approach developed in the earlier subsections.
One had better numerically simulate the hadronic activities to properly determine the source terms for the BBN equations.
Fortunately, the effects of long-lived particles heavier than 10 GeV on BBN have been explored in depth in the literature, motivated by beyond the Standard Model physics near the electroweak scale. In~\cite{Kawasaki:2004qu, Kawasaki:2017bqm}, generic BBN constraints have been set as upper bound on the primordial abundance of a new particle as a function of its lifetime, assuming it dominantly decays into specific final states (${\rm FS} = q\bar q, \tau^+\tau^-$, $e^+e^-$). Here, we reinterpret their results for the decaying heavy neutrino. Because the heavy neutrino decay mode is not unique, we apply the upper bound on $m Y$ found for a final state (FS)~\cite{Kawasaki:2017bqm} (where $Y$ is the yield of particle before its decay starts) to the quantity
\begin{equation}
   m_4 Y_{\nu_4} {\rm Br}(\nu_4\to {\rm FS} + \dots) \ ,
\end{equation}
as a function of the lifetime $\tau_4$, where $Y_{\nu_4}=n_{\nu_4}/s$ is evaluated at early times with $t\ll \tau_4$. 
If the heavy neutrino decays into more than one final state explored in~\cite{Kawasaki:2017bqm}, the BBN constraint will be the union of exclusions from each channel.
Again, we emphasize that in the simple heavy neutrino model considered here, all the quantities depend only on two fundamental parameters, $m_4$ and $|U_{a 4}|^2$.

In this work, we do not consider the heavy neutrino mass above $\sim30$ GeV for two reasons. In section~\ref{sec:2}, we show that the heavy neutrino is dominantly produced at temperatures around $T_*$. For $m_4>10$ GeV, Eq.~\eqref{eq:T*} reads (using $c_\nu=259\pi/81$ from table~\eqref{eq:GammaCA})
\begin{equation}
    T_* \simeq 80\,{\rm GeV} \left(\frac{m_4}{30
    \,\rm GeV}\right)^{1/3} \ .
\end{equation} 
Above this temperature, the neutrino scattering rate formula (Eq.~\eqref{eq:GammaA}) derived within the Fermi theory starts to breakdown. Moreover, our study focuses on early universe neutrino oscillation as the dominant production mechanism of the heavy neutrino. At temperatures above 100 GeV, the electroweak symmetry is likely restored. In that case, the active-sterile neutrino mixing shuts off, so does the oscillation. The heavy neutrino may still be produced through some Yukawa interactions but the actual production mechanism becomes sensitive to UV physics details.

\begin{figure}[t]
    \centering
    \includegraphics[width = 0.9\linewidth]{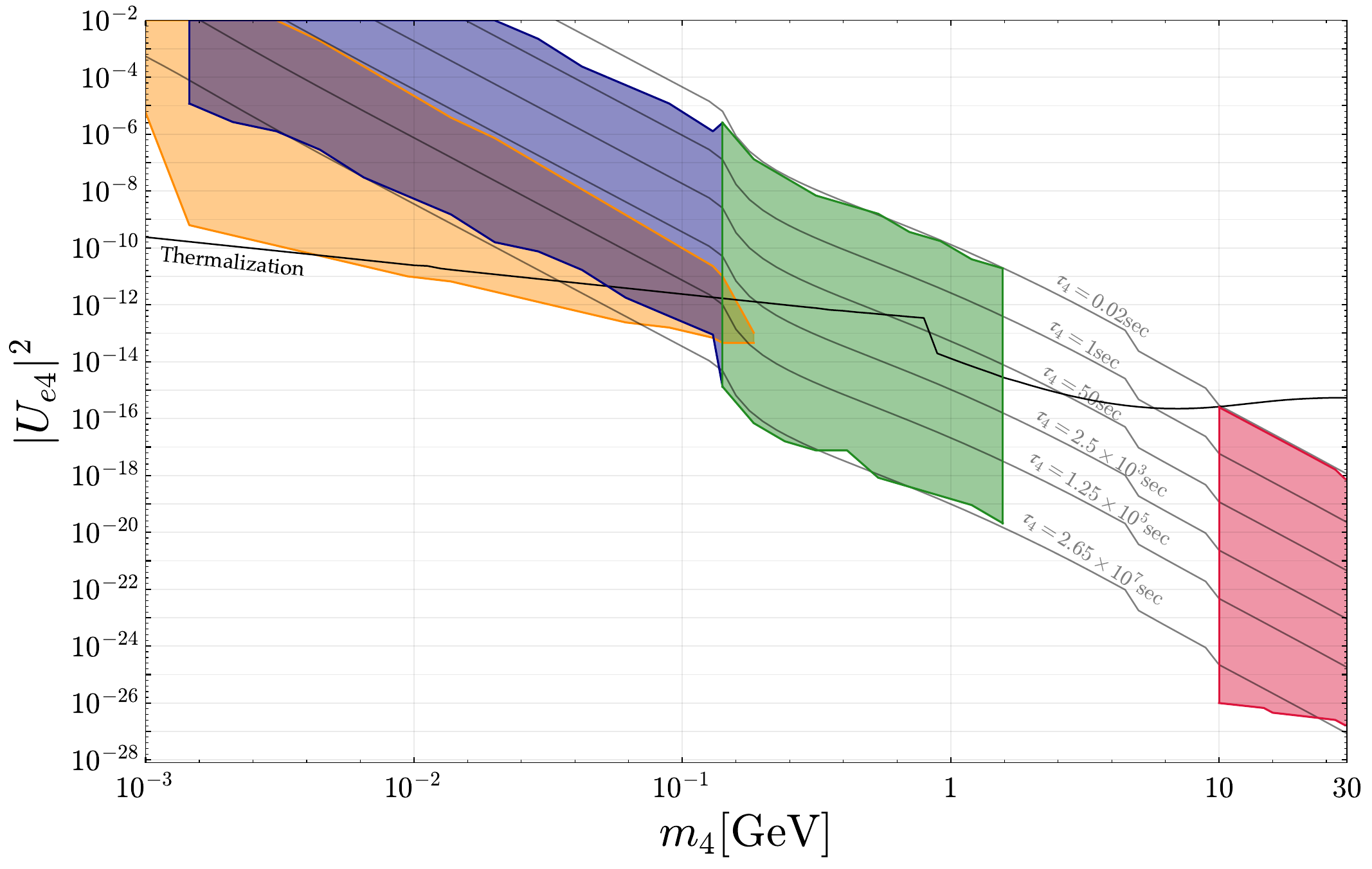}
    \caption{Anatomy of BBN constraints on the heavy neutrino mass versus mixing parameter space, with the shaded regions excluded due to various effects during BBN. Orange region is for temporary matter domination. In the blue, green and red regions, the heavy neutrino have different dominant decay modes and its decay products affect the primordial element production in different ways, detailed in the text. The black curves correspond to the thermalization boundary (see also Fig.~\ref{Fig:FOFI}), and multiple heavy neutrino lifetimes are shown.}
    \label{fig:MainPlotE2}
\end{figure}

\section{Results}

Fig.~\ref{fig:MainPlotE2} show the main result of this analysis for heavy neutrino mixing with $\nu_e$. It further divides the BBN exclusion in Fig.~\ref{Fig:MainPlotE} into various colored regions where the heavy neutrino itself or its decay products impact the BBN prediction in different ways. 

First, the effect of temporary matter domination before the decay of heavy neutrino described in section~\ref{subsec: Temporary matter domination} excludes the orange shaded region in the upper-left part of the figure, with the thermalization line penetrating through. Above this line, the heavy neutrino was once in thermal equilibrium. As discussed in section~\ref{sec:relativisitc}, the heavy neutrino always decouples relativistically. With a lifetime longer than $\sim 1$ second (corresponding to the upper boundary of the orange region), its energy density can come into domination over that of radiation during BBN. 
For heavy neutrino mass below a few MeV scale, the orange shaded region shrinks and points to larger values of $|U_{e4}|^2$ above the thermalization line. In this low mass range, a smaller $|U_{e4}|^2$ leads to too early decoupling, before the QCD phase transition, and in turn a less populous heavy neutrinos (a $g_{*S}$ effect) and weakened constraint.
In the orange shaded region below the thermalization line, the heavy neutrino never reaches thermal equilibrium with the Standard Model plasma and its abundance is built up by the freeze in mechanism described section~\ref{sec:FI}. In this region, the population of heavy neutrino is proportional to $|U_{e4}|^2$. Below the lower boundary of the orange shaded region, the freeze-in production of heavy neutrino is not efficient.

It is worth noting that the orange shaded region disappears for heavy neutrino mass above $\sim140$\,MeV. In this mass range, the heavy neutrino to pion decay channel opens up which is a two-body decay. Comparing the decay coefficients in Eqs.~\eqref{eq:LeptonicDecayCoeff} and \eqref{eq:nu4tomeson}, this gives a boost factor of $\sim(10^2-10^3)$ in the $\nu_4$'s total decay rate.
For the lifetime to remain comparable or longer than 1 second, a much smaller mixing parameter $|U_{e4}|^2$ is required. This effect manifests as the down turn of the constant lifetime ($\tau_4=1\,$sec) contours when $m_4$ goes above the pion mass threshold, and the corresponding production mechanism quickly switches from freeze out to freeze in.
As a result, the orange shaded region quickly disappears below the thermalization line where the freeze-in production is too low and temporary matter domination does not occur.

Next, we move on to discuss the effects of heavy neutrino decay on the primordial element abundances. 
We focus on deuterium and helium, which are the most produced nuclei from BBN and their abundances are the most precisely measured~\cite{ParticleDataGroup:2022pth}
    \begin{equation}\label{eq:expterror}
    \begin{aligned}
        \left. \text{D}/\text{H} \right|_\text{p} &= (25.47 \pm 0.25) \times 10^{-6} \ , \\
        \text{Y}_\text{p} &= 0.245 \pm 0.003 \ .
    \end{aligned}
    \end{equation}
To simulate predictions of the heavy neutrino model, we use the public code {\tt PRyMordial}~\cite{Burns:2023sgx} which is built with a complete network of nuclear reactions in SBBN and the portal to incorporate new physics effects. In addition to the model parameters $m_4$ and $|U_{\alpha 4}|^2, (\alpha=e,\mu,\tau)$, there are two cosmological parameters with non-negligible uncertainties, the baryon asymmetry $\eta_b$ and neutron lifetime $\tau_n$,
\begin{equation}\label{eq:inputerror}
    \begin{aligned}
    \eta_b &= 6.1 \times 10^{-10} \pm 5.5 \times 10^{-12}\ , \\
    \tau_n &= 878.4 \pm 0.5\, \text{sec} \ .
    \end{aligned}
\end{equation}
It is useful to note that the uncertainties in $\eta_b$ and $\tau_n$ propagate to the model predictions of $\left. \text{D}/\text{H} \right|_\text{p}$ and $\text{Y}_\text{p}$. The theory prediction error bars are comparable to those from experimental observations~\cite{Chowdhury:2022ahn}. We include both to derive the BBN exclusion regions.

The decay of heavy neutrino during BBN excludes the blue, green, and red shaded regions in Fig.~\ref{fig:MainPlotE2} at 95\% confidence level. 
All these exclusion regions lie below the $\tau_4 \sim 0.02$ second curve. It is worth emphasizing that this is not the absolute upper limit on the lifetime of heavy neutrino. As pointed out in the introduction, in the minimal model the BBN constraint will not apply for very long lifetime, or very small $|U_{e4}|^2$ where the heavy neutrino production is highly suppressed, leading to the lower boundary of the exclusion region. 

\begin{figure}[t]
    \centering
    \includegraphics[width = 1\linewidth]{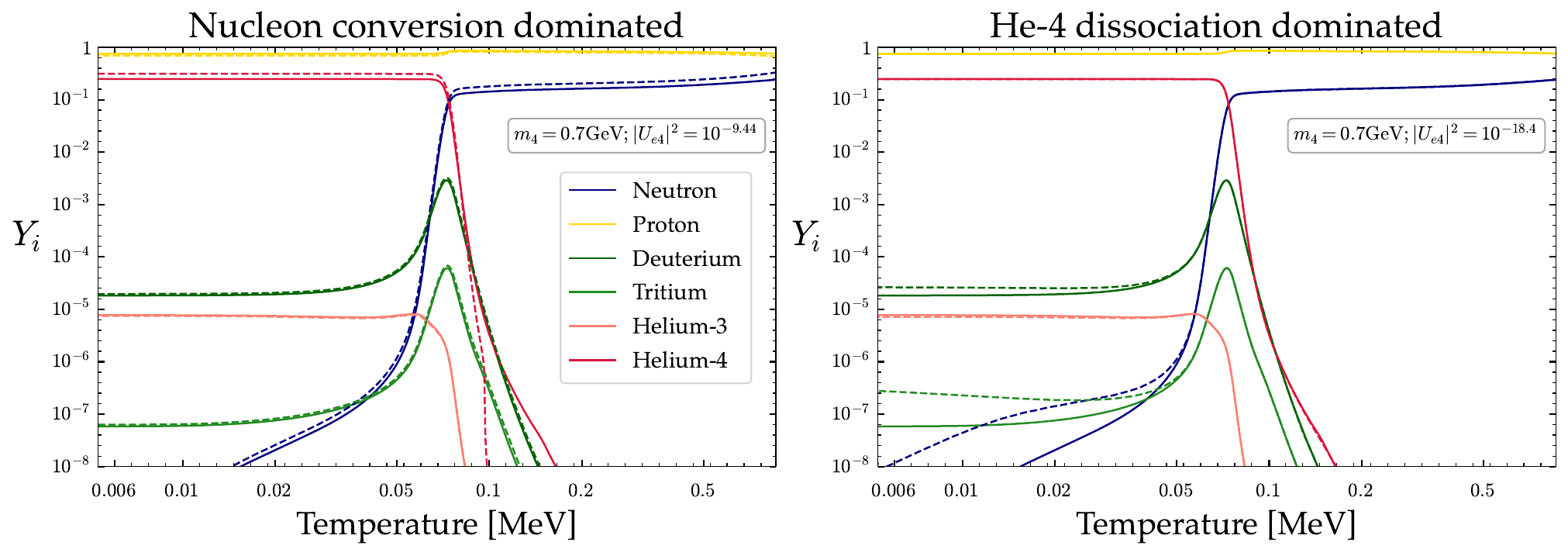}
    \caption{Temperature evolution of primordial element abundance. Time goes from right to left. Solid lines are for the standard BBN, and dashed lines are the results in the presence of a heavy neutrino that decays during BBN.
    \textbf{Left panel: } ``Large mixing'' case where the dominant BBN constraint comes from kaon and pion triggered conversion that increases the ${}^4{\rm He}$ abundance compared to standard BBN. This sets the upper edge of the BBN exclusion region in Figs.~\ref{Fig:MainPlotE} and \ref{fig:MainPlotE2}.
    \textbf{Right panel: } ``Small mixing'' case where the dominant BBN constraint is set due to hadronic ${}^4{\rm He}$ dissociation that in turn increases the deuterium abundance.
    }
    \label{fig:VaryingTemp}
\end{figure}

In the green shaded region in Fig.~\ref{fig:MainPlotE2}, with intermediate $m_4$ lying below GeV but above the meson decay threshold, we simulate kaon and pion productions using the two-body decay rates given in Eq.~\eqref{eq:nu4tomeson}.
The two most important effects in shifting the $\left. \text{D}/\text{H} \right|_\text{p}$ and $\text{Y}_\text{p}$ abundances from SBBN predictions are $pn$ conversion and hadrodissociation of nuclei triggered by $K^\pm$ and $\pi^\pm$, discussed in section~\ref{sec:chargedmeson}.
In particular, $pn$ conversion is the most important effect near the upper edge of the green shaded region which corresponds to heavy neutrino lifetime $\tau_4 \simeq0.02$ second. This value is consistent with the lifetime upper limit found in previous analyses~\cite{Boyarsky:2020dzc, Sabti:2020yrt}.
On the other hand, hadrodissociation plays the dominant role for smaller $|U_{e4}|^2$ and sets the bottom edge of the green region.
The partial decay $\nu_4\to\nu_e \pi^0$ (followed by prompt decay $\pi^0\to\gamma\gamma$) can also induce photodissociation of nuclei, discussed in section~\ref{sec:photoninj}. 
We include this effect in our analysis. Numerically, photodissociation is always subdominant to $pn$ conversion and hadrodissociation.

Fig.~\ref{fig:VaryingTemp} provides further details of our calculation by showing the temperature (time) evolution of primordial element abundances during BBN. 
The arrow of time points from right to left. 
The solid curves are SBBN predictions whereas dashed curves are for the presence of a decaying heavy neutrino.
The model parameters are chosen to be $m_4=0.7\,$GeV and $|U_{e4}|^2=10^{-9.44}$ (left), $10^{-18.4}$ (right) which correspond to a heavy neutrino lifetimes equal to $\tau_4=0.03$ and $2.65 \times 10^7$ seconds, respectively. 
In the left panel, we find that $pn$ conversion is an early effect and can be effective when the universe is only $\sim1$ sec old. The $K^\pm, \pi^\pm$ from heavy neutrino decay can increase the neutron population compared to SBBN. As a result of the higher neutron population, the so-called "deuterium-bottleneck" is widened, i.e., deuterium formation is the first step of the nucleosynthesis chain. With more deuterium, the abundance of other elements like tritium and ${}^4{\rm He}$ are also enhanced accordingly. The cost for all these increases is a deficit of the proton abundance as dashed yellow curve indicates.
The right panel shows the hadrodissociation effect. It takes place at a much later time, $t\gtrsim 100\,$sec. The injection of $K^-$ and $\pi^-$ reduces the ${}^4{\rm He}$ abundance, and in turn increases those of deuterium, tritium, and neutron.

 \begin{figure}[t]
    \centering
     \includegraphics[width = 0.618\linewidth]{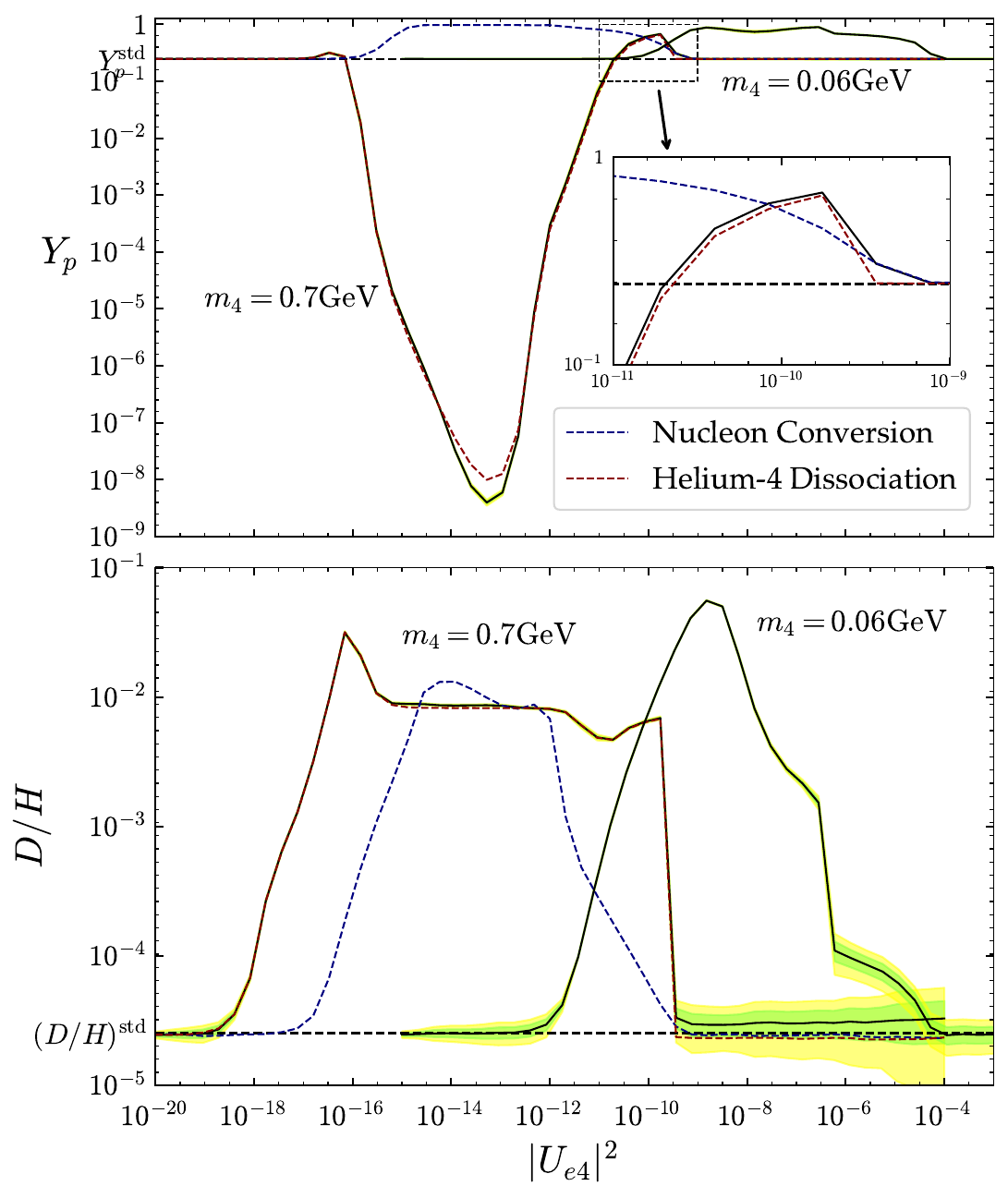}
    \caption{BBN prediction of final ${}^4{\rm He}$ (\textbf{top panel}) and deuterium (\textbf{bottom panel}) abundances as function of $|U_{e4}|^2$, for two choices of heavy neutrino mass $m_4$. In both plots, the green and regions show the 1 and $2\sigma$ theoretical uncertainty (due to uncertainties in $\eta_b$ and $\tau_n$, see Eq.~\eqref{eq:inputerror}). The central values of experimentally measured abundances are shown by black dashed lines. On each curve, there is a range of $|U_{e4}|^2$ where excessive ${}^4{\rm He}$ and deuterium are produced and is thus ruled out. 
    As discussed in the main text, for the case $m_4= 0.8$\,GeV, the dominant effects come from kaon/pion induced $pn$ conversion and ${}^4{\rm He}$ dissociation.
    The blue and red dashed curves show the results where only one of the two effects is turned on.
    }
    \label{fig:VaryingUe4}
\end{figure}

In the blue shaded region of Fig.~\ref{fig:MainPlotE2}, the $\nu_4$ mass lies below the pion threshold and has to undergo a three-body leptonic decay into $\nu e^+ e^-$ or three active neutrinos. The neutrinos in the final state can also trigger $pn$ conversion via weak interaction as discussed in section~\ref{sec:nuinj}. 
Due to the weakly interacting nature, neutrinos can free stream and build up their phase space distribution until participating in the conversion process at a later time.
We ignore the late-time nuclei dissociation effect triggered by the neutrinos or electrons which may further extend the lower boundary of the blue shaded region.
We expect this approximation will not affect our main result, because the smaller $|U_{e4}|^2$ region has already been covered by the effect of temporary matter domination, i.e., orange shaded region in Fig.~\ref{fig:MainPlotE2}.

Another useful way to view the heavy neutrino decay effects is presented in Fig.~\ref{fig:VaryingUe4}, where we hold $m_4$ fixed at two values and vary the active-sterile neutrino mixing parameter $|U_{e4}|^2$. 
The solid curves show the predictions of $\left. \text{D}/\text{H} \right|_\text{p}$ and $\text{Y}_\text{p}$ with all the new physics effects included and the brazilian bands around them stand for theoretical uncertainties in the cosmological parameters in Eq.~\eqref{eq:inputerror}.
The horizontal dashed lines show the central values of experimental measurement in Eq.~\eqref{eq:expterror}.
For each mass $m_4$, there exists a window of $|U_{e4}|^2$ where both $\left. \text{D}/\text{H} \right|_\text{p}$ and $\text{Y}_\text{p}$ significantly deviate from the SBBN predictions. For higher $m_4$, the excluded  window of $|U_{e4}|^2$ shifts to lower values. 
The far-left side to the bumps/dips, where $|U_{e4}|^2$ is very tiny, is not excluded because the heavy neutrino is not efficiently produced by the freeze-in mechanism.
The far-right side with very large $|U_{e4}|^2$ is also not excluded because the heavy neutrino decays away too early, before the onset of BBN.
The semi-analytic approach developed in this work allows us to turn on only one of the non-thermal injection effects discussed in section~\ref{sec:nu4Decay} each time. The red and blue dashed curves in Fig.~\ref{fig:VaryingUe4} shows the BBN predictions when we only take into account of the $pn$ conversion or the ${}^4{\rm He}$ hadrodissociation effect, respectively. It allows us to judge their relative importance. One can also learn that $pn$ conversion always enhances 
$\left. \text{D}/\text{H} \right|_\text{p}$ and $\text{Y}_\text{p}$, whereas hadrodissociation only enhances $\left. \text{D}/\text{H} \right|_\text{p}$ at the price of suppressing $\text{Y}_\text{p}$, consistent with Fig.~\ref{fig:VaryingTemp}.

For the high mass region with $m_4 >10$\,GeV,  we apply the existing BBN constraints in the literature as described in section~\ref{sec:largemass}. 
In this region, the exclusion extends to sufficiently small $|U_{e4}|^2$ where the heavy neutrino experiences freeze-in production in the early universe. 
Because the BBN constraint in this mass region is obtained by reinterpretation existing limits without solving the Boltzmann equations, it only provides limited information which renders us unable to trace the time evolution of element abundances, and dissect/assess the non-thermal injection effects on BBN separately, as was made possible for the lower mass regions (see Figs.~\ref{fig:VaryingTemp} and \ref{fig:VaryingUe4}).
We leave the overcoming of such limitations for a future work.

It is worth pointing out that one cannot simply apply the same constraints for the high mass region ($\gtrsim 10\,$GeV) to lighter (sub-GeV) heavy neutrinos due to the absence of baryons and antibaryons as the heavy neutrino decay product and their hadronic activities in the latter case. Otherwise, the BBN constraints would be overstated~\cite{Alonso-Alvarez:2022uxp}. 

\begin{figure}
    \centering
    \includegraphics[width = 0.9\linewidth]{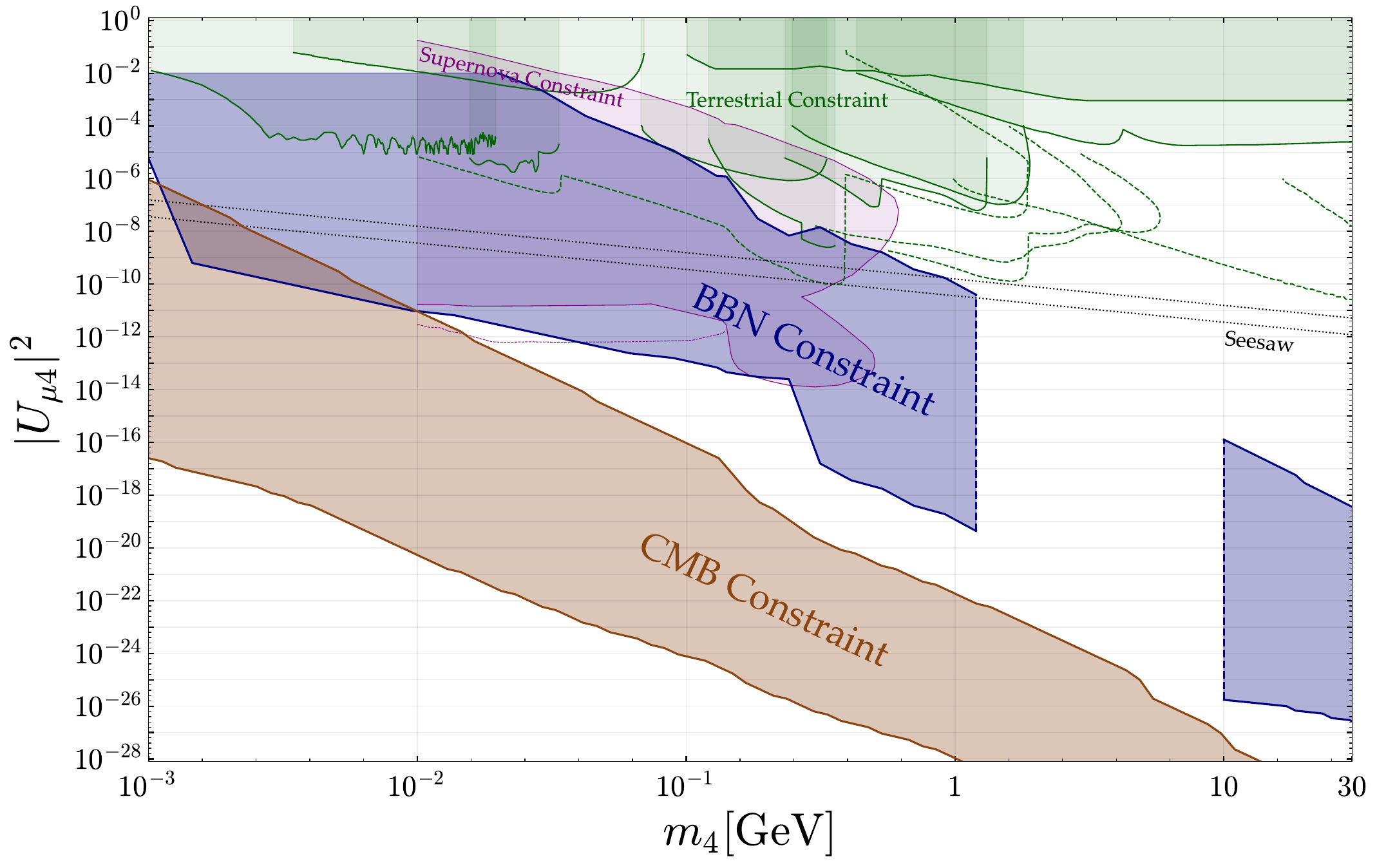}
    \includegraphics[width = 0.9\linewidth]{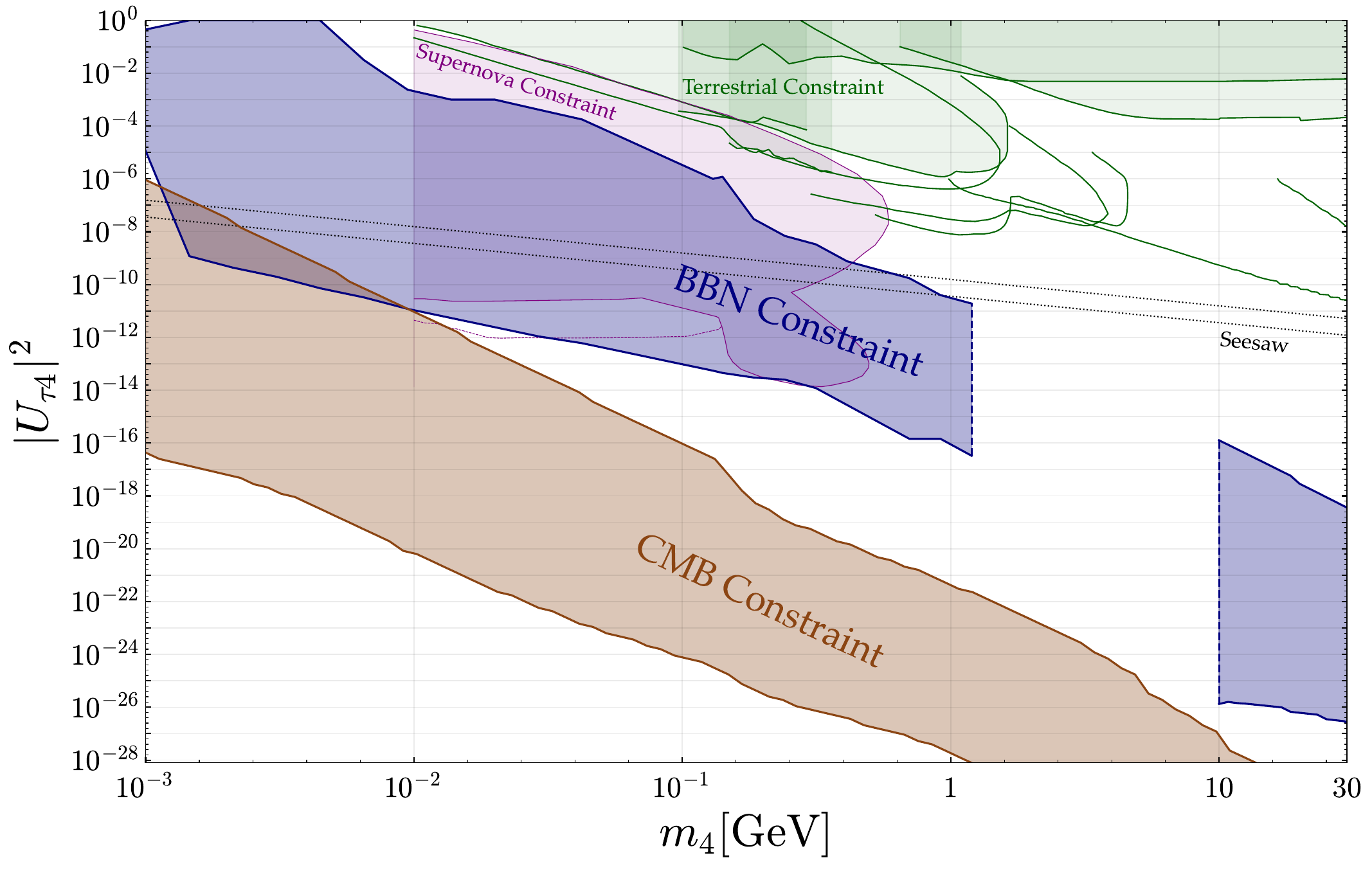}
    \caption{Similar to Fig:\ref{Fig:MainPlotE} but for heavy neutrino mixing with $\nu_\mu$ ({\bf top panel}) or $\nu_\tau$ ({\bf bottom panel}).}
    \label{Fig:MainPlotMT}
\end{figure}

For the heavy neutrino mass in the range of 1--10 GeV, one cannot analytically calculate its hadronic decays owing to the non-perturbative nature of strong interaction. Therefore, when presenting the BBN constraints in Fig.~\ref{Fig:MainPlotE} and \ref{Fig:MainPlotMT}, we leave this mass window of the heavy neutrino as a gap, understanding that the actual constraint would interpolate between those derived for the lower and higher mass regions.

Finally, we also explore the heavy neutrino mixing with $\nu_\mu$ or $\nu_\tau$ flavors in a similar fashion to $\nu_e$. Their results are presented in Fig.~\ref{Fig:MainPlotMT}. The physics is similar, except for the modification of mass windows due to the non-negligible muon and tau lepton masses. 
We also include a constraint derived using core-collapse supernova~\cite{Carenza:2023old, DelaTorreLuque:2024zsr, Akita:2023iwq}.
Compared to previous works, our results present the entire BBN excluded region and a panoramic view over the heavy neutrino parameter space.

\section{CMB Constraint on very Long-lived Heavy Neutrino}

For heavy neutrino with lifetime much longer than the BBN timescale, we consider another useful constraint on the energy injection into the universe during the formation of CMB. 
We adopt the CMB constraint result set on decaying dark matter previously derived in~\cite{Slatyer:2016qyl}, which puts an lower bound on the lifetime on a particle $X$ assuming it comprises 100\% of dark matter in the universe, i.e., $\Omega_c h^2=0.12$.
To translate it for the long-lived heavy neutrino, we note that the population is not tied to $\Omega_c$ but determined by the freeze-out or freeze-in mechanisms, as discussed in Section~\ref{sec:2}. In this case, the lifetime lower limit is set for
\begin{equation}\label{eq:cmb1}
  \tau_4 > [\tau_X]_{\rm min} \times\left[\frac{n_4\left(m_4, |U_{a 4}|^2\right)}{n_X}\right]_{T=0.3\,\rm eV} \ ,
\end{equation}
where $n_4$ is the number density of heavy neutrino $\nu_4$ evolved to the time of recombination, where $T\simeq 0.3$\,eV, and it includes the exponential decaying factor.
$n_X \simeq (T/2.7\,{\rm K})^3\rho_c \Omega_c/m_X$ is the dark matter number density at the same time and $\rho_c=1.05\times10^{-5}h^2\,{\rm GeV/cm^3}$ is the critical density of today's universe.
$[\tau_X]_{\rm min}$ as a function of $m_X$ (set to be equal to $m_4$ here) is obtained by digitalizing Fig.~8 of~\cite{Slatyer:2016qyl} (approximately, $[\tau_X]_{\rm min}\simeq 10^{25}\,$s for $m_X$ above MeV scale). The limit from Eq.~\eqref{eq:cmb1} translates into the lower boundary of the brown shaded region in Figs.~\ref{Fig:MainPlotE} and \ref{Fig:MainPlotMT}. Near this lower boundary, the heavy neutrino has a lifetime longer than the age of the universe~\cite{Alonso-Alvarez:2022uxp} and it can be part of the dark matter in the universe. As a useful comparison, the CMB constraint on $|U_{a4}|^2$ is stronger than the indirect detection limit on $\nu_4\to\nu\gamma$ decay found in~\cite{Essig:2013goa}.

Similar to the argument made for BBN, the CMB constraint does not apply for sufficiently large $|U_{a4}|^2$ where the heavy neutrino decay happens too early. Without detailed analysis, we simply set this upper boundary as where the heavy neutrino lifetime is equal to the age of universe when $T\simeq 10\,$eV, which corresponds to the onset of large scale structure formation~\cite{Kolb:1990vq}.
There is a gap of allowed parameter space between the BBN and CMB exclusion regions.

\section{Conclusion}

In this work, we explore the BBN constraints on a heavy neutrino in the minimal model. The heavy neutrino is introduced as a pure gauge singlet fermion and it could interact with other known particles only through mixing with active neutrinos after the electroweak symmetry breaking. Such a setup is found in beyond the Standard Model theories of the type-I seesaw mechanism for neutrino mass generation, or neutrino portal dark sectors. When the mixing occurs with a single active neutrino flavor $\nu_\alpha, \ (\alpha=e,\mu,\tau)$, the minimal has only two new parameters, the heavy neutrino mass $m_4$ and the mixing $|U_{a4}|^2$, defined in Eq.~\eqref{eq:DefMixing}. We point out that not only the decays of the heavy neutrino are fully determined by the two parameters, but also its production mechanism in the early universe. For sufficiently large $|U_{a4}|^2$, the relic abundance of heavy neutrino (before decaying away) is set by the thermal freeze-out mechanism similar to the decoupling of light neutrinos. For smaller $|U_{a4}|^2$, the production takes place via active-sterile neutrino oscillation in the early universe plasma (a kind of freeze-in mechanism). For each point in the $m_4$ versus $|U_{a4}|^2$ parameter space, we calculate the heavy neutrino's abundance using either of the above production mechanism and feed the result as the input for our BBN analysis. It is useful to point out that the time scales for the heavy neutrino production and decay of heavy neutrino are always well separated.

For the BBN constraints, we consider the effects of temporary matter domination prior to the decay of heavy neutrino, as well as non-thermal energy injections due to the decay.
For the latter case, we develop a semi-analytic approach to derive the BBN constraints for heavy neutrino mass below the GeV scale.
We find that hadronic energy injections, in the form of $K^\pm, \pi^\pm$ and through the $pn$ conversion and nuclei hadrodissociation processes, plays the dominant role. 
The injection of light neutrinos from the decay is important for heavy neutrino mass below $\sim100$ MeV.
BBN can robustly constrain the parameter space of heavy neutrino mass $m_4$ from MeV to the weak scale.
It is sensitive to very small mixing parameter $|U_{a4}|^2$ where the heavy neutrino lifetime is as long as $\sim 10^7$ seconds.
For even smaller mixings, the BBN constraint eventually runs out the steam because the heavy neutrino cannot be abundantly produced via the freeze in mechanism.
In that region, we sketch a CMB constraint on the very long-lived heavy neutrino using the energy injection argument.

Our analysis offers a complete description of the fate of an unstable heavy neutrino and its impact on the early universe. Its production and decay are both determined by the two minimal model parameters, with little room for assumptions. For the freeze-in production mechanism, we assume zero initial abundance of the heavy neutrino. This assumption could be relaxed with additional new physics at high scale, but they would only result in an enhanced heavy neutrino abundance and stronger BBN and CMB constraints. In contrast, the freeze-out mechanism is not sensitive to high scale physics and the BBN constraint in that region of parameter space is robust. 

Our main results, Figs.~\ref{Fig:MainPlotE} and \ref{Fig:MainPlotMT}, show that BBN excludes the low-scale type-I seesaw mechanism with heavy neutrino mass below a few hundred MeV. By combining with other existing constraints, this conclusion holds even in the presence of the caveat for large $|U_{a4}|^2$ noticed in~\cite{Casas:2001sr, Kersten:2007vk}.
BBN and CMB are at the frontier of probing very weakly-coupled heavy neutrinos and complementary to the role of terrestrial high-energy and/or high-intensity experiments.

\section*{Acknowledgment}

We thank Anne-Katherine Burns, Tammi Chowdhury, James Cline, André de Gouvêa, Seyda Ipek, Miha Nemev\v{s}ek, Tim Tait, Douglas Tuckler, Mauro Valli for helpful discussions. 
We thank Kazunori Kohri for valuable communications on BBN constraints in the large mass region and baryon (anti-baryon) injection effects.
This work is supported by a Subatomic Physics Discovery Grant (individual) from the Natural Sciences and Engineering Research Council of Canada.
YMC thanks the Galileo Galilei Institute for Theoretical Physics for hospitality and partial support while this work was under preparation.

\appendix
\section{Boltzmann equation for active-sterile neutrino oscillation}\label{app:A}

To describe the collisional oscillation for dark matter production, we consider the phase-space density matrix in the $(\nu_a, \nu_s)$ space
\begin{equation}
\mathcal{F} = \begin{pmatrix}
f_{a} & f_{as} \\
f_{sa} & f_{s} 
\end{pmatrix} \ ,
\end{equation}
where the diagonal elements $f_a (f_s)$ stand for phase space distributions for the active (sterile) neutrino. Quantum effects lie in the off-diagonal elements.

The density matrix evolves according to the quantum kinetic equations~\cite{Sigl:1993ctk, Chu:2006ua, Hannestad:2013ana},
\begin{equation}\label{eq:kineticequation}
\begin{split}
H \frac{\partial f_{a}}{\partial \log a} &=  \frac{i}{2} \Delta \sin(2\theta) \left(f_{as} - f_{sa}\rule{0mm}{4mm}\right) + \Gamma_A \left(f_{a} - f_{\rm eq}\rule{0mm}{4mm}\right) \ , \\
H \frac{\partial f_{as}}{\partial \log a} &=  \frac{i}{2}\left[ \Delta \sin(2\theta) \left(f_{a}-f_{s} \rule{0mm}{4mm}\right) + 2 \left(\Delta \cos(2\theta)-V \rule{0mm}{4mm}\right) f_{as} \rule{0mm}{6mm}\right] - \frac{\Gamma}{2} f_{as}  \ , \\
H \frac{\partial f_{sa}}{\partial \log a} &=  - \frac{i}{2}\left[ \Delta \sin(2\theta) \left(f_{a}-f_{s}\rule{0mm}{4mm}\right) + 2 \left(\Delta \cos(2\theta)-V\rule{0mm}{4mm}\right) f_{sa} \rule{0mm}{6mm}\right] - \frac{\Gamma}{2} f_{sa}  \ , \\
H \frac{\partial f_{s}}{\partial \log a} &= - \frac{i}{2} \Delta  \sin(2\theta) \left(f_{as} - f_{sa}\rule{0mm}{4mm}\right) \ . \\
\end{split}
\end{equation}
In the first equation, $\Gamma_A$ is relevant for the active neutrino's chemical potential and equals to their total annihilation/co-annihilation rate.
On the other hand, $\Gamma$ in the second and third equation is relevant for the decoherence of the neutrino state and receives contribution from both annihilation and scattering rates. It is given by Eq.~\eqref{eq:GammaA} and \eqref{eq:GammaCA}.
$\Gamma$ must come with a prefactor $1/2$ because it multiplies with the off-diagonal elements of the density matrix (wavefunction is the ``square root'' of density matrix).
All the other quantities involved in this equation are also defined below Eq.~\eqref{eq:PSDBoltzmannEq}. 

To proceed, we consider early universe where the temperature is much higher than MeV and the weak interaction rates $\Gamma, \Gamma_A$ are much higher than the Hubble parameter $H$.
This bring about two simplifications to the set of equation in Eq.~\eqref{eq:kineticequation}. First, $\Gamma_A\gg H$ locks $f_a$ for active neutrinos to the thermal distribution
\begin{equation}
f_a = f_{\rm eq} \ .
\end{equation}
With this, the second equation of Eq.~\eqref{eq:kineticequation} can be written as
\begin{equation}\label{eq:A4}
\frac{\partial f_{as}}{\partial \log a} =
\left[  \frac{i}{H} \left(\Delta \cos(2\theta)-V\rule{0mm}{4mm}\right)- \frac{\Gamma}{2 H}  \rule{0mm}{6mm}\right]  f_{as} + \frac{i}{2H}\Delta \sin(2\theta) \left(f_{\rm eq} - f_s\rule{0mm}{4mm}\right)  \ .
\end{equation}
For clarify, we define $\tau=\log a$,
\begin{equation}
\begin{split}
g(\tau) &\equiv \frac{i}{H} \left(\Delta \cos(2\theta)-V\rule{0mm}{4mm}\right)- \frac{\Gamma}{2 H} \ , \\
h(\tau) &\equiv \frac{i}{2H}\Delta \sin(2\theta) \left(f_{\rm eq} - f_s\rule{0mm}{4mm}\right) \ , 
\end{split}
\end{equation}
and rewrite Eq.~\eqref{eq:A4} as
\begin{equation}
\frac{d f_{as}}{d\tau} = g(\tau) f_{as} + h(\tau) \ .
\end{equation}
With initial condition $f_{as}(0)=0$, the solution to the non-homogeneous equation takes the form
\begin{equation}\label{eq:A7}
f_{as}(\tau) = \int_0^\tau d \tau' h(\tau') \exp\int_{\tau'}^\tau d\tau'' g(\tau'') \ .
\end{equation}

In the limit $\Gamma \gg H$, we have $\left|{\rm Re}[g(\tau)]\right|\gg1$ but ${\rm Re}[g(\tau)]<0$. As a result, the $\tau'$ integral in the above equation is dominated by contributions with $\tau'\lesssim \tau$, otherwise the exponential factor is strongly suppressed. This observation allows us to write
\begin{equation}\label{eq:A8}
f_{as}(\tau) \simeq h(\tau) \int_0^\tau d \tau'  \exp\int_{\tau'}^\tau d\tau'' g(\tau'') \simeq \frac{h(\tau)}{g(\tau)} \int_0^\tau d \tau' g(\tau') \exp\int_{\tau'}^\tau d\tau'' g(\tau'') \ .
\end{equation}

Next we introduce 
\begin{equation}
G(\tau', \tau) \equiv \int_{\tau'}^\tau d\tau'' g(\tau'') \quad \Rightarrow \quad  dG(\tau', \tau) = - g(\tau') d\tau' \ ,
\end{equation}
which allows Eq.~\eqref{eq:A8} to be rewritten as
\begin{equation}\label{eq:A10}
\begin{split}
f_{as}(\tau)  &\simeq - \frac{h(\tau)}{g(\tau)} \int_0^\tau  e^{G(\tau', \tau)} d G(\tau', \tau) \\
&= - \frac{h(\tau)}{g(\tau)} \exp\left[G(\tau', \tau)\right]^{\tau'=\tau}_{\tau'=0}  \\
&=- \frac{h(\tau)}{g(\tau)} \left[1 - \exp\int_0^\tau d\tau' g(\tau') \right] \\
&\simeq - \frac{h(\tau)}{g(\tau)} \ .
\end{split}
\end{equation}
In the last step, we again used that the factor $-\Gamma/(2H)$ in $g(\tau)$ causes a strong damping to the exponential term allowing it to be dropped.

With the above simplification, we derive
\begin{equation}
f_{as}= f_{sa}^* \simeq
 \frac{i\Delta \sin(2\theta)/2}{ \Gamma/2 -  i (\Delta \cos(2\theta) - V)} \left(f_{\rm eq} - f_s\rule{0mm}{4mm}\right)  \ .
\end{equation}
Plugging these results into the last equation of Eq.~\eqref{eq:kineticequation}, we finally obtain
\begin{equation}
\begin{split}
\frac{\partial f_{s}}{\partial \log a} 
&\simeq \frac{\Gamma}{4H} \left[ \frac{\Delta^2 \sin^2(2\theta)}{(\Delta\cos(2\theta) -V)^2  + \Gamma^2/4} \right] \left(f_{\rm eq} - f_s\rule{0mm}{4mm}\right) \ .
\end{split}
\end{equation}
This agrees with the Boltzmann equation Eq.~\eqref{eq:PSDBoltzmannEq} in the small $\theta$ limit. 
In Eq.~\eqref{eq:PSDBoltzmannEq}, we added a $\Delta^2 \sin^2(2\theta)$ term to the denominator inside the square bracket, which makes it equal to $\sin^2(2\theta_{\rm eff})$ defined in Eq.~\eqref{eq:thetaeff} and correctly approach to $\sin^2(2\theta)$ in the low temperature limit ($\Gamma, V\to0$).

\section{Heavy neutrino decay rates in the large mass region}\label{app:B}

In this appendix, we present the decay rate of heavy neutrino with mass close to $W, Z$ boson masses. In this case, we use the gauge boson propagators in calculation instead of the Fermi theory. We continue assume to that the heavy neutrino contains a small mixture of active neutrino flavor $\nu_a$.
For simplicity, we take all final state fermions to be massless.

For the heavy neutrino $\nu_4$ lighter than $W$ or $Z$, we present the coefficients $\tilde c$ for each decay rate as defined in Eq.~\eqref{eq: decay width}. 
\begin{itemize}
\item $\nu_4$ decays via  off-shell $W$ exchange (only one diagram)
\begin{equation}
\begin{split}
\tilde c_{\nu_4 \to u_i \bar d _j \ell^-_a} &= \left(\frac{|V_{ij}^{\rm CKM}|^2}{64}\right) \times 12 \int_0^1 dx \int_0^{1-x} dy \frac{x(1-x)\Omega_W^2}{(y-\Omega_W)^2 + \zeta_W^2} \ , \\
\tilde c_{\nu_4 \to \nu_b \ell^+_b \ell^-_a} &= \left(\frac{1}{192}\right) \times 12 \int_0^1 dx \int_0^{1-x} dy \frac{x(1-x)\Omega_W^2}{(y-\Omega_W)^2+ \zeta_W^2} \ , 
\end{split}
\end{equation}

\item $\nu_4$ decays via off-shell $Z$ exchange (only one diagram)
\begin{equation}
\begin{split}
\tilde c_{\nu_4 \to \nu_a \nu_b \bar{\nu}_b} &= \left(\frac{1}{768}\right) \times 12 \int_0^1 dx \int_0^{1-x} dy \frac{x(1-x)\Omega_Z^2}{(y-\Omega_Z)^2 + \zeta_Z^2} \ , \\
\tilde c_{\nu_4 \to \nu_a u_i \bar{u}_i} &= \left(\frac{5}{2304}\right) \times \frac{12}{5} \int_0^1 dx \int_0^{1-x} dy \frac{(5x+y-5x^2-y^2-2x y)\Omega_Z^2}{(y-\Omega_Z)^2+ \zeta_Z^2} \ , \\
\tilde c_{\nu_4 \to \nu_a d_i \bar{d}_i} &= \left(\frac{13}{4608}\right) \times \frac{6}{13} \int_0^1 dx \int_0^{1-x} dy \frac{(26x+y-26x^2-y^2-2x y)\Omega_Z^2}{(y-\Omega_Z)^2+ \zeta_Z^2} \ , \\
\tilde c_{\nu_4 \to \nu_a \ell^+_b \ell^-_b} &= \left(\frac{1}{1536}\right) \times 6 \int_0^1 dx \int_0^{1-x} dy \frac{(2x+y-2x^2-y^2-2x y)\Omega_Z^2}{(y-\Omega_Z)^2+ \zeta_Z^2} \ , \\
\end{split}
\end{equation}

\item $\nu_4$ decay into three neutrinos (two diagrams interferencing with each other, both with off-shell $Z$ exchange)
\begin{equation}
\begin{split}
\tilde c_{\nu_4 \to \nu_a \nu_a \bar{\nu}_a} &= \left(\frac{1}{384}\right) \times 3 \int_0^1 dx \int_0^{1-x} dy \frac{x(1-x)\Omega_Z^2[(x-1+2\Omega_Z)^2+4\zeta_Z^2]}{[(y-\Omega_Z)^2+\zeta_Z^2][(x+y-1+\Omega_Z)^2+ \zeta_Z^2]} \ .
\end{split}
\end{equation}

\item $\nu_4 \to \nu_a \ell_a^-\ell_a^+$ decay (two diagrams interferenceing with each other, one with off-shell $W$ and the other with off-shell $Z$ exchange)
\begin{equation}
\begin{split}
\tilde c_{\nu_4 \to \nu_a \ell^+_a \ell^-_a} &\simeq \left(\frac{5}{1536}\right)\times \frac{6}{5} \int_0^1 dx \int_0^{1-x} dy \left\{\frac{(2x-2x^2-2xy+y-y^2)\Omega_Z^2}{(y-\Omega_Z)^2 + \zeta_Z^2}
\right.\\
&\hspace{0cm}\left. + \frac{16x(1-x)\Omega_W^2}{(x+y+\Omega_W-1)^2 + \zeta_W^2} + \frac{8x(1-x)\Omega_W \Omega_Z(y-\Omega_Z)(x+y+\Omega_W-1)}{[(y-\Omega_Z)^2+\zeta_Z^2][(x+y+\Omega_W-1)^2+\zeta_W^2]} \right\} \ .
\end{split}
\end{equation}
\end{itemize}
In the above formulae, we define $\Omega_W=M_W^2/m_4^2$, $\zeta_W=M_W \Gamma_W/m_4^2$, $\Omega_Z=M_Z^2/m_4^2$ and $\zeta_Z=M_Z \Gamma_Z/m_4^2$. In the limit $\Omega, \Omega_W\to \infty$, the above $\tilde c$ coefficients returns to the simple forms given in Eqs.~\eqref{eq:LeptonicDecayCoeff} and \eqref{eq:nu4toquark}.

If the heavy neutrino is heavier than $W, Z$, it undergoes a two-body decay. The corresponding decay rates are
\begin{equation}
\begin{split}
\Gamma_{\nu_4 \to \ell_a^-W^+} &=  \frac{G_F m_4^3 |U_{a4}|^2 }{8\sqrt2 \pi}
\left( 1 - \frac{3 M_W^4}{m_4} + \frac{2 M_W^6}{m_4^3}\right) \ , \\
\Gamma_{\nu_4 \to \nu_a Z} &=  \frac{G_F m_4^3 |U_{a4}|^2 }{16\sqrt2 \pi}
\left( 1 - \frac{3 M_Z^4}{m_4} + \frac{2 M_Z^6}{m_4^3}\right)
\end{split}
\end{equation}
It is worth pointing out the longitudinal enhancement in the limit of $m_4\gg M_{W,Z}$ as a consequence of the Goldstone equivalence theorem.

\section{Photodissociation cross sections}\label{app:C}

In our analysis, we implement the photodissociation processes for elements up to ${}^4{\rm He}$.
The corresponding cross sections are the numerical fits taken from the appendix of~\cite{Cyburt:2002uv}.

\begin{enumerate}
    \item $d ( \gamma, n) p, \hspace{1em} E_{\gamma, \text{th}} = \left| Q \right|= \SI{2.224573}{\MeV}$ \\ [1.5ex]
    $\sigma (E_\gamma) = \SI{18.75}{\milli \barn}$
    $\left[ \left( \frac{\sqrt{\left| Q \right|(E_\gamma - \left| Q \right|}}{E_\gamma} \right)^3 + 0.007947 \left( \frac{\sqrt{\left| Q \right|(E_\gamma - \left| Q \right|}}{E_\gamma} \right)^2 \frac{\left( \sqrt{\left| Q \right|}-\sqrt{0.037} \right)^2}{E_\gamma - \left( \left| Q \right| - 0.037 \right)}\right] $

    \item $t ( \gamma, n) d, \hspace{1em} E_{\gamma, \text{th}} = \left| Q \right|= \SI{6.257248}{\MeV}$, \hspace{0.5em}
    $\sigma (E_\gamma) = \SI{9.8}{\milli \barn}$
    $\frac{\left| Q \right|^{1.95} \left( E_\gamma - \left| Q \right| \right)^{1.65}}{E_\gamma^{3.6}}$

    \item $t ( \gamma, np) n, \hspace{1em} E_{\gamma, \text{th}} = \left| Q \right|= \SI{8.481821}{\MeV}$, \hspace{0.5em}
    $\sigma (E_\gamma) = \SI{26.0}{\milli \barn}$
    $\frac{\left| Q \right|^{2.6} \left( E_\gamma - \left| Q \right| \right)^{2.3}}{E_\gamma^{4.9}}$

    \item $^3He ( \gamma, p) d, \hspace{1em} E_{\gamma, \text{th}} = \left| Q \right|= \SI{5.493485}{\MeV}$, \hspace{0.5em}
    $\sigma (E_\gamma) = \SI{8.88}{\milli \barn}$
    $\frac{\left| Q \right|^{1.75} \left( E_\gamma - \left| Q \right| \right)^{1.65}}{E_\gamma^{3.4}}$

    \item $^3He ( \gamma, np) p, \hspace{1em} E_{\gamma, \text{th}} = \left| Q \right|= \SI{7.718058}{\MeV}$, \hspace{0.5em}
    $\sigma (E_\gamma) = \SI{16.7}{\milli \barn}$
    $\frac{\left| Q \right|^{1.95} \left( E_\gamma - \left| Q \right| \right)^{2.3}}{E_\gamma^{4.25}}$

    \item $^4He ( \gamma, p) t, \hspace{1em} E_{\gamma, \text{th}} = \left| Q \right|= \SI{19.813852}{\MeV}$, \hspace{0.5em}
    $\sigma (E_\gamma) = \SI{19.5}{\milli \barn}$
    $\frac{\left| Q \right|^{3.5} \left( E_\gamma - \left| Q \right| \right)^{1.0}}{E_\gamma^{4.5}}$

    \item $^4He ( \gamma, n) ^3He, \hspace{1em} E_{\gamma, \text{th}} = \left| Q \right|= \SI{20.577615}{\MeV}$, \hspace{0.5em}
    $\sigma (E_\gamma) = \SI{17.1}{\milli \barn}$
    $\frac{\left| Q \right|^{3.5} \left( E_\gamma - \left| Q \right| \right)^{1.0}}{E_\gamma^{4.5}}$

    \item $^4He ( \gamma, d) d, \hspace{1em} E_{\gamma, \text{th}} = \left| Q \right|= \SI{23.846527}{\MeV}$, \hspace{0.5em}
    $\sigma (E_\gamma) = \SI{10.7}{\milli \barn}$
    $\frac{\left| Q \right|^{10.2} \left( E_\gamma - \left| Q \right| \right)^{3.4}}{E_\gamma^{13.6}}$

    \item $^4He ( \gamma, np) d, \hspace{1em} E_{\gamma, \text{th}} = \left| Q \right|= \SI{26.0711}{\MeV}$, \hspace{0.5em}
    $\sigma (E_\gamma) = \SI{21.7}{\milli \barn}$
    $\frac{\left| Q \right|^{4.0} \left( E_\gamma - \left| Q \right| \right)^{3.0}}{E_\gamma^{7.0}}$
\end{enumerate}

\bibliography{BBNReferences}
\bibliographystyle{JHEP}

\end{document}